\documentclass[aps,10pt,prd,a4paper,twocolumn,notitlepage,nofootinbib,floatfix]{revtex4-1}
\usepackage{amsmath}
\usepackage{graphicx}
\usepackage{color}
\usepackage[normalem]{ulem} 

\allowdisplaybreaks

\newcommand{\be}{\begin{equation}}
\newcommand{\ee}{\end{equation}}

\newcommand{\pa}{\partial}

\newcommand{\bea}{\begin{eqnarray}}
\newcommand{\eea}{\end{eqnarray}}
\newcommand{\ben}{\begin{eqnarray*}}
\newcommand{\een}{\end{eqnarray*}}

\begin{document}

\title{Rotating systems, universal features in dragging and anti-dragging effects, and bounds onto angular momentum}
 
\author{Janusz Karkowski}
\author{Patryk Mach}\email{patryk.mach@uj.edu.pl}
\author{Edward Malec}\email{malec@th.if.uj.edu.pl}
\author{Micha\l~ Pir\'og}
\affiliation{Instytut Fizyki im.~Mariana Smoluchowskiego, Uniwersytet Jagiello\'nski, {\L}ojasiewicza 11, 30-348 Krak\'{o}w, Poland} 
\author{Naqing Xie}\email{nqxie@fudan.edu.cn}
  \affiliation{School of Mathematical Sciences, Fudan University, Shanghai, China}
 
\renewcommand{\labelitemii}{$\cdot$}
\renewcommand{\labelitemiii}{$\diamond$}
\renewcommand{\labelitemiv}{$\ast$}

\renewcommand{\labelitemi}{$\bullet$}
\renewcommand{\[}{\begin{equation*}}
\renewcommand{\]}{\end{equation*}}

\begin{abstract}
We consider stationary, axially symmetric toroids rotating around spinless black holes, assuming the general-relativistic Keplerian rotation law, in the first post-Newtonian approximation. Numerical investigation shows that the angular momentum accumulates almost exclusively within toroids. It appears that various types of dragging (anti-dragging) effects are positively correlated with the ratio $M_\mathrm{D}/m$ ($M_\mathrm{D}$ is the mass of a toroid and $m$ is the mass of the black hole) --- moreover, their maxima are proportional to $M_\mathrm{D}/m$. The horizontal sizes of investigated toroids range from c.\ 50 to c.\ 450 of Schwarzschild radii $R_\mathrm{S}$ of the central black hole; their mass $M_\mathrm{D} \in (10^{-4}m, 40m)$ and the radial size of the system is c.\ 500 $R_\mathrm{S}$. We found that the relative strength of various dragging (anti-dragging) effects does not change with the mass ratio, but it depends on the size of toroids. 

Several isoperimetric inequalities   involving angular momentum  are shown to hold true.
\end{abstract}

\maketitle
\section{Introduction}
\label{sec1}

There are three  principal aims of this paper. 

 We have found recently two new weak field effects that affect angular velocities of gaseous disks   rotating around spinless black holes \cite{JMMP, MM2015}. They appear in the first post-Newtonian approximation 
 (1PN hereafter),  in addition to the well known geometric dragging of frames.  One of them --- we call it anti-dragging, since it works against the   dragging of frames --- is proportional to the speed of sound of gas.
  The other depends on a combination of gravitational and centrifugal potentials, that strictly vanishes for weightless disks. All 1PN corrections  strictly vanish for uniformly rotating disks, but they are 
  nonzero for the important case of the Keplerian rotation. 
 
 We shall  address in this paper  the following question: what are the principal physical properties of a (Keplerian) rotating toroidal-black-hole system, 
 that are responsible for the strength of various dragging (anti-dragging) effects?  One   can expect --- by appealing to the behaviour of test particles in the Kerr geometry --- that
    robust dragging phenomena should be  associated with compact systems that possess a lot of  angular momentum.  Our investigation shows that this intuition is incorrect, and that the relevant characteristic is  the mass ratio  $M_\mathrm{D}/m$, where $M_\mathrm{D}$ is the  mass of a  toroid and $m$ is the mass of the black hole.  We find an interesting universality: the maxima of the combined, normalized in a suitable sense,
(1PN) corrections, as well of its constituents --- the  geometric dragging, the anti-dragging, and the  centrifugal one \cite{MM2015} --- are simply proportional to $M_\mathrm{D}/m$, for a fixed extension of a toroid.  We have studied polytropic disks for two classes of polytropes; the universality appears in all examples, but some numerical coefficients depend (albeit   rather weakly) on   the equation of state of the fluid.

An interesting question is the influence of a rotating environment onto the central black hole. There are reports  --- for  a  rigid rotation   \cite{nishida_eriguchi, Ansorg_Petroff}  and the constant specific angular momentum \cite{shibata} --- that the black hole can carry substantial amount of the angular momentum. We 
assume a general-relativistic version of the Keplerian rotation law \cite{MM2015}. It comes as a surprise, that one can have compact systems with a large amount of angular momentum,  where central black holes practically do not participate in rotation ---  their spin parameters  are smaller than $10^{-4}$, and they carry less than one-millionth of the total angular momentum.
 
Finally, there exist several inequalities that must be obeyed by quasilocal characteristics of apparent horizons. S. Dain extended these by formulating local estimates onto local angular momentum of rotating bodies, and proved them, under somewhat stringent conditions \cite{Dain, Daingrg}.   We show that they  are in fact satisfied.

The order of the rest of this paper is as follows. In the next section we formulate axial perturbations of
the conformal Schwarzschild geometry, that describe toroids rotating around a spinless black hole. The basic idea is to build systems such that the gravity of toroids is negligible --- compared to the gravity of the central black hole --- close to the event horizon,
and the gravitational potential is small within the bulk of rotating matter. The rotation is ruled by the general-relativistic Keplerian law.  Section \ref{sec3} shows the first post-Newtonian approximation to equations of motion. Section \ref{sec4} is dedicated to the quasilocal description of the system, in particular to defining the concept of the apparent horizon, its mass and its angular momentum. We briefly describe the essence of the numerical method in Sec.\ \ref{sec5}. Section \ref{sec6} brings   main numerical results concerning various 1PN corrections to  the angular velocity; we emphasise again the unexpected universality of these effects. Section \ref{sec7} addresses the issue of distribution of the angular momentum. It appears that the black hole spin parameter $a_\mathrm{S}\equiv cj_\mathrm{S}/m^2$ is very small; the central black hole is almost Schwarzschildean.  Only a tiny fraction of the total angular momentum can be attributed to the black hole; this is commented and explained  therein. In Sec.\ \ref{sec8} we review Dain's results on estimations of the angular momentum. Tables I and II allow one to find out that the angular momentum can be bounded, as postulated in \cite{Dain}. The last section summarizes obtained results.

We assume throughout the paper the gravitational constant $G=1$.   There is a scaling freedom that allows us to treat the speed of light $c$ as a free parameter.  We adjust the speed of light $c$ and the coordinate extension of the disk, in Secs.\ \ref{sec6}--\ref{sec8}, so that the whole system has an (approximate) areal size of $500 R_\mathrm{S}$, and the inner edges of the investigated disks are located between $50 R_\mathrm{S}$ -- $475 R_\mathrm{S}$ (again approximately).  The total asymptotic mass $M_\mathrm{D}+m$ ranges between $m$ and $40m$. We take care to  construct numerical solutions that are in the 1PN regime within the stationary toroids.  

\section{Axial  perturbations  of the conformal Schwarzschild geometry}
\label{sec2}
 
Einstein equations, with the signature of the metric $(-,+,+,+)$, read
\begin{equation}
R_{\mu \nu} -g_{\mu \nu }\frac{R}{2} =  \frac{8\pi }{c^4}T_{\mu \nu },
\label{ee1}
\end{equation}
where $T_{\mu \nu }$ is the stress-momentum tensor.  The  metric is given by

\begin{align}
\label{metric}
d s^2 &=  -  e^{\frac{2\nu }{c^2} }(d x^0)^2
+r^2   e^{\frac{2\beta }{c^2} } \left( d \phi  -  \frac{A_\phi  }{r^2c^3} d x^0\right)^2
\nonumber\\&\qquad
+e^{\frac{2 \alpha }{c^2} }   \left( dr^2 +dz^2\right)     .
\end{align}
Here $x^0 =ct$ is the rescaled  time coordinate, and $r$, $\phi$, $z$ are cylindrical coordinates.  
 We assume axial and equatorial symmetry, and stationarity --- thus metric functions $\nu , \alpha , \beta $ and $A_\phi  $ depend only or $r$ and $z $ ---  and employ the stress-momentum tensor  of the perfect fluid 
\be 
\label{emD}
T^{\alpha\beta} = \rho (c^2+h)u^\alpha u^\beta + p g^{\alpha\beta},
\ee
where $\rho$ is the baryonic rest-mass density, $h$ is the  specific enthalpy,
and $p$ is the  pressure.  The Greek indices  range from 0 to 3, and the Latin indices change from 1 to 3.
The 4-velocity  {$u^\alpha =\frac{dx^\alpha }{cd\tau }$} along the world line of fluid particles  (here $\tau $ is their proper time)  is normalized: $g_{\alpha\beta}u^\alpha u^\beta=-1$. We introduce $u^t\equiv u^0/c$.
 The coordinate (angular)  velocity of the fluid reads ${\vec v}= \Omega \partial_\phi $, where $\Omega = u^\phi /u^t$. 

We assume the polytropic  equation of state   $p(\rho ,s) = K(s) \rho^\gamma$,
where $s$ is the specific entropy of fluid and $\gamma$ is a constant. Then one has $h(\rho ,s) = K(s) \frac{\gamma}{\gamma-1}\rho^{\gamma-1}$. The entropy is assumed to be constant.

We shall  study small \emph{stationary}, cylindrically symmetric,  perturbations of the Schwarzschild spacetime, with  the angular momentum  being carried by a rotating   disk of fluid. 
We  consider  the  following  geometry in the conformal (cylindrical) coordinates,
\begin{eqnarray}
 ds^2&= &-(dx^0)^{2}  \left( \frac{f_+}{f_-}\right)^2  +      \left(f_-\right)^4       \left( dr^2+ dz^2 +  r^2d\phi^2 \right) \nonumber\\
&&
-2 \frac{A_\phi }{c^3} \left( r, z \right)      d \phi  d x^0  ,
\label{sc}
\end{eqnarray}   
where the two functions $f_+$ and $f_-$ are defined  as
\begin{eqnarray}
f_+&=&1+\frac{U}{2c^2}, \nonumber\\
f_-&=&1-\frac{U}{2c^2}.
\label{ff}
\end{eqnarray}
Here the gravitational potential $U$  is a superposition of the central term ($-\frac{m}{R}$) and $U_\mathrm{D}$, induced by the disk:
\begin{equation}
U= -\frac{m}{R}+U_\mathrm{D}.
\label{U}
\end{equation}
Henceforth $R=\sqrt{r^2+z^2}$. 

Let  us point out  that in the metric (\ref{sc})  the  lapse function    $ N\equiv   \frac{f_+}{f_{-}}$  and   the shift vector $X_i=(0,0,-A_\phi /c^3 )$ are $\phi$-independent.

We shall say that  perturbations are small, if    $|U_\mathrm{D}|/c^2 \ll1$. In addition, we require that within the volume $V$  of a rotating disk $m/(Rc^2)\ll 1$; the two facts imply $|U|/c^2 \ll 1$ inside a disk. Then it is legitimate to perform the approximation procedure --- the expansion in powers of $1/c^2$. It appears that the  line element (\ref{sc}) leads to stationary equations that coincide, in the  1PN approximation, with stationary equations corresponding to the  1PN approximation of the metric (\ref{metric}). \textit{Thus the two approaches are equivalent up to the  1PN order. }

     The case, when the disk's potential $U_\mathrm{D}$ and the metric function   $A_\phi  $ do vanish, yields  strictly the Schwarzschild line element. 
In this case  the parameter $m$ is just the asymptotic  mass, the event and apparent horizons coincide and they are located at $R_\mathrm{h}=m/(2c^2)$. We should point that the metric (\ref{sc}) is more convenient than (\ref{metric}), because  it is easier to describe horizons of  black holes in the class of conformal   deformations of the Schwarzschild metric, than in the general metric (\ref{metric}). This description can be regarded as a version of the effective field approximation.   Another application of the conformal factorization of the metric can be found in \cite{Shibata97}, where post-Newtonian equilibria of co-rotating neutron star binaries are investigated. 
 
It is well known that equations of the stationary Einstein hydrodynamics are not closed; this is similar to the stationary Newtonian hydrodynamics.  One needs to impose an additional closure assumption --- a general-relativistic version of the ``rotation curve'' known  in the Newtonian hydrodynamics --- in order to complete the system.  In the  Newtonian case one defines directly the rotation curve --- the angular velocity $\Omega $ --- as a prescribed function of the distance from the rotation axis. If the $z$-axis is the symmetry axis, then $\Omega =\Omega (r )$.

 Rotation curves --- angular velocities as functions of coordinates ---  emerge in the context of  general relativity   as solutions of the following three-step procedure. We define $\omega \equiv r^{-2} A_\phi$.
First, notice that  
\be
j  \equiv c^2 u_\phi u^t= \frac{v^2}{\left( \Omega -\frac{\omega }{c^2}\right) \left( 1-\frac{v^2}{c^2}\right) },
\ee
where 
\be
v^2 \equiv r^2 \left( \Omega -\frac{\omega }{c^2}\right)^2 e^{ -4U/c^2},
\ee
can be interpreted as the angular momentum per unit  inertial  mass   \cite{Moncrief}.
 The resulting system   is integrable  if $j$ depends only on the angular velocity $\Omega $, $ j\equiv j(\Omega)$ \cite{ Bardeen_1973, Bardeen_1970, Butterworth_Ipser, nishida1}. In the second step one needs to specify  $j(\Omega )$, in order to close the description of a stationary system.    There are few  options in the existing literature, from the simple linear function  $j(\Omega )=A(\Omega  -B)$, where $A$ and $B$ are constants  \cite{ Bardeen_1973, nishida1}, to recent nonlinear proposals of \cite{GYE, MM2015, UTGHSTY}.  We adopt, following \cite{MM2015}, the rotation law
\be
\label{momentum}
j(\Omega ) \equiv \frac{w^{\frac{4}{3}}  \Omega^{-\frac{1}{3}} }{1+  \frac{ 3 }{   c^2 }  w^{\frac{4}{3} }\Omega^{ \frac{2}{3} } +\frac{4\Psi }{   c^2}  }.
  \ee
Here $w$ and $\Psi $ are   parameters that shall be determined at each step of the post-Newtonian expansion.  This choice yields  the Keplerian rotation law   in the Newtonian limit \cite{MM2015}; therefore we sometimes refer to (\ref{momentum}) as the ``general-relativistic Keplerian rotation law''.  

The rotation curves  $\Omega \left( r, z \right) $ can now be recovered --- in the final step --- from  the equation 
\be
\label{rotation_law}
 \frac{w^{\frac{4}{3}}  \Omega^{-\frac{1}{3}} }{1+  \frac{ 3 }{   c^2 }  w^{\frac{4}{3} }\Omega^{ \frac{2}{3} } +\frac{4\Psi }{   c^2}  }  = \frac{v^2}{\left( \Omega -\frac{\omega }{c^2}\right) \left( 1-\frac{v^2}{c^2}\right) }.  
\ee

 We would like to note that the rotation law (\ref{momentum}) leads to a new prediction, that was absent in \cite{JMMP}, namely to the centrifugal correction $-\frac{3}{2c^2} \Omega_0 \left( -\frac{m}{\tilde r}+ U_{0D}+\Omega_0^2\tilde r^2\right) $ to the angular velocity --- see Eq. (\ref{angular velocityc2}) in the next section,  and  explanations therein. It is interesting also that the weightless disk of dust rotating according to  (\ref{rotation_law}) satisfies exactly the Einstein equations within the Schwarzschild spacetime \cite{MM2015}.  

\section{1PN equations within a rotating disk}
\label{sec3}

The   1PN equations for rotating fluids have been derived  in \cite{JMMP}  and \cite{MM2015}, on the basis of an earlier work \cite{BDS}. In this section we present a brief description of the reasoning.   
  
The  metric becomes, assuming $|U|\ll c^2$,
\begin{align}
\label{metric1}
ds^2 &=  -  \left(1+\frac{2U}{c^2} +\frac{2U^2}{c^4}\right) (d x^0)^2
-2  c^{-3} A_\phi d x^0 d \phi \ 
\nonumber\\&\qquad
+\left(1-\frac{2U}{c^2}  \right)  \left( d r^2 + d z^2 + r^2 d \phi^2\right)    ;
\end{align} 
its spatial part is conformally flat.

The 1PN approximation  can be valid if
  $|U|\ll c^2$. 
 
 We split different quantities  ($\rho$, $p$, $h$, $U  $ and $v^i $) into their Newtonian (denoted by subscript 0) and 1PN (denoted by subscript 1) parts.
This splitting reads, for  the specific enthalpy $h$,  the density $\rho $, the angular velocity $\Omega$, the quantity  $\Psi $     and   the  potential $U$:  
\begin{subequations}
\label{density_rotation}
\begin{align}
h&=h_0+c^{-2} h_1, \\[1ex]
\rho &= \rho_0 + c^{-2} \rho_1,
\\[1ex]
 \qquad
\Omega  &=\Omega_0  + c^{-2} v_1^\phi , 
\\[1ex] 
\qquad
\Psi  &= \Psi_0  + c^{-2}  \Psi_1 ,
\\[1ex] 
\qquad
U& = -\frac{m}{R}+U_{0D} + c^{-2} U_1.
\end{align}
\end{subequations}

The angular velocity becomes Keplerian,  
\begin{equation}
\Omega_0=\frac{w}{r^{3/2}},
\end{equation}
in the Newtonian limit. Let  $ M_\mathrm{0D}=\int_VdV \rho_0$, where $dV$ is the geometric volume element,  denote the Newtonian mass of the disk.
It appears from numerical analysis that the parameter $w$ is close to $\sqrt{m}$, if $M_\mathrm{0D}\ll m$. We expect (basing on numerics and partial analytical results) 
that in general $\sqrt{m}\le w\le \sqrt{m+M_\mathrm{0D}} $ \cite{MMP2013}.

Notice that, up to the 1PN order,   
\begin{equation}
\label{enthalpy}
\frac{1}{\rho} \partial_i p = \partial_i h_0 + c^{-2} \partial_ih_1
 {+ \mathcal{O}(c^{-4})},
\end{equation}
where the {1PN} correction $h_1$ to the specific enthalpy can be written as $h_1 = \frac{dh_0}{d\rho_0} \rho_1$. For the polytropic equation of state this gives $h_1= \left( \gamma -1 \right) h_0 \rho_1 / \rho_0$.
 
One can obtain the  Bernoulli equations at the  0PN and 1PN orders, using the  foregoing splitting into Newtonian (0PN) and 1PN parts.
 
The 0PN Bernoulli equation reads
\be
\label{0Bernoulli}
h_0-\frac{m}{R}+U_{0D}+ \Omega^2_0r^2  = \Psi_0,
\ee
where $\Psi_0$ is a constant that  can be interpreted as the binding energy per unit mass.
At the Newtonian level this is supplemented by the Poisson equation for the gravitational potential 
\be
\label{DeltaU0}
\Delta U_{0D} = 4\pi   \rho_0,
\ee
where $\Delta$ denotes the flat Laplacian.  

The component $A_\phi $ of the shift vector  satisfies the following equation
\be
\label{Afi}
\Delta A_\phi -2\frac{\pa_rA_\phi }{r}= -16 \pi  r^2 \rho_0 \Omega_0 .
\ee

The  1PN   correction  $U_1$ to the potential is obtained by solving the equation
\begin{equation}
\label{DeltaU1}
\Delta U_1 = 4\pi  \left[   \rho_1 + 2p_0 + \rho_0(h_0-2U_0+2 r^2 \Omega_0^2) \right].
\end{equation}
The mass of a disk, including the 1PN correction, is given by the volume integral 
\begin{eqnarray}
M_\mathrm{D}&=&\int_V dV   \frac{1}{c^2}\left[  \rho_1 + 2p_0 + \rho_0(h_0-2U_0+2 r^2\Omega_0^2 ) \right]
\nonumber\\
&&+M_\mathrm{0D}. 
\label{mass}
\end{eqnarray}

Finally, we have  the Bernoulli equation of the  1PN order,
\begin{eqnarray}
\Psi_1 & = & -h_1 - U_1 -\Omega_0 A_\phi + 2 r^2 \Omega_0^2 h_0 - \frac{3}{2} h^2_0 \nonumber \\
& & - 4 h_0 U_0 - 2 U_0^2 -  4 r^4\Omega^4_0,
\label{cc}
\end{eqnarray}
where $\Psi_1$ is a constant.

     In vacuum  we are left with a pair of  homogeneous elliptic equations 
\begin{eqnarray}
\label{DeltaU}
&&\Delta U  = 0,\nonumber\\
&&\Delta A_\phi -2\frac{\pa_r A_\phi }{r}= 0.
\end{eqnarray}

 The system of equations (\ref{density_rotation}--\ref{cc}) fully describes disk configurations 
up to the  1PN order of approximation. It follows from the metric (\ref{metric}) or (\ref{sc}); as we stressed before, both approaches give the same set of equations.
 Notice that these are integro-differential equations with a free boundary; the shape of a disk comes  as a part of the solution. It  will be explained   in the numerical section, how to deal with such a boundary problem. The second of the equations (\ref{DeltaU})  can cause difficulties for systems with matter located on the $z$ axis; we deal, however, with disks and Eq.\ (\ref{DeltaU}) is harmless.

The first post-Newtonian correction $v^\phi_1$ to the angular velocity $\Omega $ is obtained from the perturbation expansion of the rotation law (\ref{rotation_law}) up to terms of the order $c^{-2}$. 
 One arrives at \cite{MM2015}
 \begin{equation}
\label{constraint_solution_th}
v^\phi_1 = - \frac{3}{2} \Omega_0^3 r^2 + \frac{3A_\phi}{4 r^2 } - 3\Omega_0 h_0.
\end{equation}
 
This can be written in a more geometric way. The geometric circumferential radius of the circle $r=\mathrm{const  }$, $z = 0$ around  the rotation axis is given by $\tilde r=r  (1-U_0/c^2)+\mathcal{O}(c^{-4})$.    Therefore   the angular velocity reads
 \begin{eqnarray}
\label{angular velocityc2}
\Omega &=&\Omega_0+\frac{v^\phi_1}{c^2} =\frac{w}{\tilde r^{3/2}}    -\frac{3}{2c^2} \Omega_0 \left( -\frac{m}{\tilde r}+ U_{0D}+\Omega_0^2\tilde r^2\right) \nonumber\\
&&\ + \frac{3A_\phi}{4\tilde r^2 c^2}   -  \frac{3}{c^2}  \Omega_0 h_0, 
\end{eqnarray}
where  the   last three terms are 1PN corrections. Let us point out that the anti-dragging term ($ -  \frac{3}{c^2}  \Omega_0 h_0 $) is of the opposite sign to the dragging term  $\frac{3A_\phi}{4\tilde r^2 c^2}$ \cite{JMMP}.

\section{Characteristics of the black hole horizon }
\label{sec4}

The event horizon  coincides, in the stationary case, with the apparent  horizon $S$ \cite{Wald}, that is a two-surface $S$ embedded in the 3-hypersurface $\Sigma $ defined by $t = \mathrm{const}$ such that
\begin{eqnarray}
\nabla_in^i +h_{ij}K^{ij}&=&0, \nonumber\\
 \nabla_in^i -h_{ij}K^{ij}&>&0.
\label{ah}
\end{eqnarray}
Here $n^i$ is the unit normal to the surface $S$, $X_i$ denotes the shift vector, $K_{ij}=\frac{1}{2N}\left( \nabla_i X_j+ \nabla_iX_{j}\right) $   is the  extrinsic curvature of $\Sigma $, $h_{ij} =g_{ij}-n_in_j$ is the induced metric on $S$ and $\nabla_i $ means the covariant derivative on $\Sigma $. In our case the shift vector is given by $X_i=(0,0, -A_\phi/c^3)$. It is easily seen that for the metric (\ref{sc}) the contribution of extrinsic curvature vanishes (albeit some nondiagonal components of $K_{ij}$ are nonzero); thus the  apparent horizon becomes  a minimal surface. The three-dimensional metric of $\Sigma $ in 
spherical coordinates is given by 

\begin{eqnarray}
 ds^2_3= 
  \left(f_-\right)^4       \left( dR^2+ R^2d\theta^2 +  R^2\sin ^2\theta d\phi^2 \right)  ,
\label{sc2}
\end{eqnarray}   
 and we get from  the minimal surface equation within $\Sigma $
 \begin{equation}
\nabla_in^i=0.
\label{min}
\end{equation}
The two-surface $S$  can be assumed  to be $\phi$-independent, due to the axial symmetry; it is given by   the single equation $R=R(\theta )$. Its normal reads
\begin{equation}
\label{wektor}
n_k=\frac{f_-^2}{\sqrt{1+\frac{(\partial_\theta R)^2}{R^2}}} (1, -\partial_\theta R, 0).
\end{equation}
In explicit terms Eq.\ (\ref{min}) reads
\begin{eqnarray}
&& \frac{1}{R^2f_-^6}\partial_R \left[  \frac{R^2f^4_- }{\sqrt{1+\frac{(\partial_\theta R)^2}{R^2}}}\right] +
\nonumber\\
&& \frac{1}{\sin \theta R^2f_-^6\ }  \partial_\theta  
\left[  \frac{\sin \theta (-\partial_\theta R) f^4_-}{\sqrt{1+\frac{(\partial_\theta R)^2}{R^2}}}\right] = 0.
\label{minimal}
\end{eqnarray}
 
Equation (\ref{minimal}) is a second-order differential equation for the radial function $R(\theta )$. The boundary conditions are given by $\partial_\theta R|_{\theta =0}=\partial_\theta R|_{\theta =\pi }=0$. Solving of Eq. (\ref{minimal}) would be numerically inexpedient, since the 3-metric $g_{ij}$ is known only numerically.  Fortunately, numerical results show that  the impact of  the disk onto the geometry around $R = R_\mathrm{h}$  is small.   We investigated the behaviour  of $U_\mathrm{D}$ at $R_\mathrm{h}$ for 4 disk configurations with the outer edge at $r_\mathrm{out} = 1$ and with the inner edges located at  $r_\mathrm{in}=0.1, 0.5, 0.75$ and  $ r_\mathrm{in}=0.95$. Disk masses  exceeded the central mass by a factor of 10. We found that the modulus of the potential $U_\mathrm{D}$ depends very weakly on the angle $\theta $ and its maximal values are   24.82   (for $r_\mathrm{in}=0.1$), 13.45 (for $r_\mathrm{in}=0.5$),  11.49 (for $ r_\mathrm{in}=0.75$) and 
   10.41 (for $ r_\mathrm{in}=0.95$). (See one of later sections for a more detailed description of assumed data.) These values should be compared to the modulus of the central potential $m/R_\mathrm{h}$, which gives 2000 (we assumed $m=1$ in these numerical calculations), that is  roughly   values larger by 2 orders of magnitude.
Thus  the potential $U_\mathrm{D}$  is relatively small and it is almost constant at surfaces $R=\mathrm{const}$; moreover $|\partial_RU_\mathrm{D}|/(m/R^2)\approx 0$, in the vicinity of $R_\mathrm{h}=m/(2c^2)$. 

 We believe, basing on the preliminary inspection of   Eq. (\ref{minimal}), that the following hypothesis is true.

\textit{Conjecture.} If the disk potential and its  derivatives   are small, 
\begin{eqnarray}
&&\sup_V\left(  |U_\mathrm{D}| , |R\partial_RU_\mathrm{D} | \right) \ll \frac{m}{2R}, \nonumber\\
&&
\sup_V\left(  |\partial_\theta U_\mathrm{D}|, |\partial_\theta^2 U_\mathrm{D}|\right) \ll \frac{m}{R},
\label{c1}
\end{eqnarray}
in a vicinity of $R_\mathrm{h} =m/(2c^2)$, then the position of the minimal surface is well approximated by $R_\mathrm{h} $. 

   This Conjecture will be applied later on.   

\subsection{The area and mass of minimal surfaces}

 The metric induced on an axially symmetric  two-surface $S$,  that is defined by   $R=R(\theta )$, reads
   \begin{eqnarray}
&&ds^2_{(2)}=  f_-^4   R^2 \left[  \left( 1+ \frac{(\partial_\theta R)^2}{R^2}\right) d\theta^2 +\sin^2\theta d\phi^2 \right] . 
\label{induced_metric}
\end{eqnarray}   
The area of the two-surface is given by

  \begin{equation}
 A_\mathrm{AH}=2\pi \int_0^\pi d\theta R^2 \sin \theta f_-^4\sqrt{1+\frac{(\partial_\theta R)^2}{R^2}} .
 \label{min1}
 \end{equation} 
  If the 2-surface is  minimal and it is  located at $R\approx m/(2c^2)$, then the area is approximated by   
 
 \begin{equation}
 A_\mathrm{AH}=\frac{m^2}{2c^4}\pi \int_0^\pi d\theta   \sin \theta  \left[ 16+\frac{16U_\mathrm{D}(R,\theta )}{c^2} \right].
 \label{min2}
 \end{equation}
In the case of the Schwarzschild black hole one has the strict equality $A_\mathrm{AH}= 16\pi \frac{m^2}{c^4}$. The Podurets-Misner-Hawking-Geroch \cite{Podurets} mass of a black hole is defined as $M_\mathrm{AH}=c^2\sqrt{\frac{A_\mathrm{AH}}{16\pi }}$, which yields $M_\mathrm{AH}=m$ for the Schwarzschild black hole.  It is   accepted in numerical general relativity as a quasilocal mass measure of horizons  in selfgravitating systems \cite{Szabados}. We shall also employ that  mass.  We will accept only  those numerical solutions, where the Podurets-Misner-Hawking-Geroch    mass is close to the Newtonian mass $m$, that is  $|m-M_\mathrm{AH}(S)|\ll m$ (see Sec.\ \ref{sec6}). In such a case the mass parameter $m$ still can be interpreted as the  mass of the central black hole, and it would be legitimate to use the above Conjecture.

In the  1PN approximation   the asymptotic mass of configurations  can be read off from the asymptotic behaviour of the superposition of  potentials  $-\frac{m}{R}+U_\mathrm{D}$.   As explained above,  the parameter $m$ approximates the mass of the central black hole; therefore  the asymptotics of $U_\mathrm{D}$ can be regarded as defining  the mass of a rotating disk.

 \subsection{The angular momentum    }
 
Cylindrically symmetric stationary systems possess  a (spatial) Killing vector $\eta^\mu $,  i.e. $\nabla_\mu \eta_\nu +\nabla_\nu \eta_\mu=0$.
 The angular momentum of a  fluid rotating around the $z$-axis can be defined as follows \cite{Moncrief}
\begin{eqnarray}
L & = &\frac{1}{c}\int_V t_\mu T^\mu_\nu \eta^\nu dV \nonumber \\
& = & \int_V drd\phi dz \rho (c^2 +h )  \sqrt{-\det g^4} u_\phi u^t.
\label{volume_ang}
\end{eqnarray}
We use here cylindrical coordinates $r, \phi$ and $z$; $t^\mu$ is a normal to  the Cauchy hypersurface $\Sigma $.
The integrand can be expressed as a total divergence (using  relevant Einstein equations);
applying the Gauss theorem, one can write down the angular momentum as a sum of internal angular momenta, associated to an internal 2-surface $S$ and to the asymptotics  \cite{Bardeen_1973, nishida2}:
$L=j_\mathrm{S} +j_\infty $.

The angular momentum of a surface $S$, that is  defined by $R=R(\theta )$, reads (up to  the 1PN approximation)
 
 \begin{eqnarray}
 j_\mathrm{S}&=&  \int_0^\pi  d\theta \frac{ R^4}{8} \sin^3\theta   \left( -\partial_R\omega + \frac{ \partial_\theta R \partial_\theta \omega }{R^2}\right)  \times \nonumber\\
 && \frac{f_-^7}  {   |f_+|     } ,
\label{kret} 
\end{eqnarray} 
while the asymptotic angular momentum is given by
 \begin{eqnarray}
 j_\infty &=&  -\int_0^\pi  \frac{d\theta }{8} R^4 \sin^3\theta   \left( -\partial_R\omega  \right)   .
\label{kretinfinity} 
\end{eqnarray}
  We shall calculate the angular momentum of a minimal surface.
 Assuming that the above   conjecture is valid,  that is $R\approx m/(2c^2)$, then $ (f_-)^7\approx 2^7(1-7U_\mathrm{D}/(4c^2))=128(1-7U_\mathrm{D}/(4c^2))$ and  $ |f_-| \approx |U_\mathrm{D}|/(2c^2)$.   Thus the angular momentum becomes
  
 \begin{eqnarray}
&& j_\mathrm{S}= -16\int_0^\pi  d\theta  R^4 \sin^3\theta  \left( \partial_R\omega -\frac{1}{R^2} \partial_\theta R \partial_\theta \omega \right)  \times 
\nonumber\\ &&
 \frac{1-7\frac{U_\mathrm{D}(r,\theta )}{4c^2} }  {\frac{|U_\mathrm{D}|}{c^2}}.
 \label{kret1}
 \end{eqnarray} 
 If the angular dependence of the minimal surface is ``moderate'', so that terms with $\partial_\theta R$ can be neglected, then  taking into account that    $|U_\mathrm{D}|/c^2\ll 1$,  we get
 \begin{eqnarray}
&& j_\mathrm{S}= -16\int_0^\pi  d\theta   R^4 \sin^3\theta  \frac{ \partial_R\omega }{\frac{|U_\mathrm{D}|}{c^2}}.
 \label{kret2}
 \end{eqnarray}
This can be further simplified, since $R\approx m/(2c^2)$.  We obtain finally 
  \begin{eqnarray}
&& j_\mathrm{S}= -\frac{m^4}{c^6} \int_0^\pi  d\theta    \frac{  \sin^3\theta  \partial_R\omega }{|U_\mathrm{D}|}.
 \label{kret3}
 \end{eqnarray} 
The total angular momentum of a rotating fluid is approximated, up to the 1PN order, by 

\begin{eqnarray}
L&= &\int d r d\phi dz    \Omega_0 r^3\rho_0   +
\nonumber\\ &&
\frac{1}{c^2}\int  dr d\phi dz  \rho_0 \left(    \Omega_0^3 r^5 -6  \Omega_0 r^3 U+r^3v^\phi_1 -rA_\phi  \right) 
 \nonumber\\ &&
 + \frac{1}{c^2}\int  dr d\phi dz (\rho_1 +p_0)  r^3  \Omega_0 .
\label{am}
\end{eqnarray}
 The asymptotic value of the angular momentum can be obtained directly or (preferably) from  $ j_\infty =L-j_\mathrm{S}$.

\section{Description of the numerical method}
\label{sec5}

The numerical method used in this paper was described in \cite{JMMP}. It is a variant of the old fashioned (but working) Self-Consistent Field (SCF) scheme. The solution representing the disk is found in three stages. In the first one, a strictly Newtonian configuration is obtained. In this stage, each iteration of the SCF method consists in solving the Poisson equation (16) for the gravitational potential and, subsequently, the Euler-Bernoulli equation (15) that yields the specific enthalpy corresponding to a given gravitational potential. The Newtonian solution is parametrized by the values $r_\mathrm{in}$ and  $r_\mathrm{out}$ of the inner and outer equatorial radii of the disk, respectively --- we fix the \textit{coordinate size of the disk}. Further data include the maximum value of the density within the disk, the value of the central mass $m$, and the polytropic exponent. These parameters allow one to establish the values of the constants $K$, $\Psi_0$, and $w$ in each of the subsequent iterations.

In the second stage we compute the potential $A_\phi$, solving Eq.\ (17). The result depends on previously obtained $\rho_0$ and $\Omega_0$.

The third stage is again an iterative one. In each iteration, a post-Newtonian correction $U_1$ is found by solving Eq.\ (18). Given the new value of $U_1$ we obtain the correction to the specific enthalpy $h_1$ from Eq.\ (20). There is a degree of freedom in this step that is connected with the choice of constant $\Psi_1$ in Eq.\ (20). At the Newtonian stage the analogous constant $\Psi_0$ is fixed by the geometrical requirement on the size of the disk (setting the inner and outer equatorial radii) and the remaining data (the maximum value of the density, the value of the central mass $m$, and the polytropic exponent). It seems that at the first post-Newtonian level one can choose the constant $\Psi_1$ freely. In \cite{JMMP} the value of $\Psi_1$ was fixed by requiring that the correction to the specific enthalpy $h_1$ vanishes at the outermost point of the disk, i.e., at the outer equatorial radius. We found that this choice yields solutions that conform with the post-Newtonian assumptions only in a limited range of parameters. The choice that we employ here is to require that the post-Newtonian correction $h_1$ (or, equivalently, $\rho_1$) vanishes at the point where the Newtonian density $\rho_0$ attains its maximum value. It seems to be  a slight change, but it yields acceptable solutions for a much broader range of parameters. \textit{This fact probably conveys an important message,  that in this kind of general-relativistic free boundary problems, that we are dealing with, the maximal (baryonic) mass density should always be a part of given data.} {\textcolor{red}{}}Note also that the apparent freedom of choosing the value of the constant $\Psi_1$ is yet another manifestation of the non-uniqueness of the post-Newtonian expansion with respect to stationary systems \cite{MM2015}.

 In summary, we adopted following input data: the coordinates of the inner and outer boundaries of the disk, the maximal mass density, and a {\it functional form} of the rotation law and 
of the equation of state. That is, we assume $p=K\rho^\gamma $, but the specific value of the coefficient $K$ is a part of a solution. Similarly, the parameters $w$ and $\Psi $  in the rotation law  --- see Eq. (\ref{rotation_law}) --- are established after finding the Newtonian (or 1PN) disk configuration. 

Technical aspects of our numerical method are quite standard; they are described in \cite{JMMP}. We work in spherical coordinates $(R, \theta, \phi)$. For convenience we also define $\mu = \cos \theta$. Equations (16), (17) and (18) are solved using appropriate Green functions that are expanded in Legendre polynomials. 

Equations (16) and (18) have the form of a Poisson equation
$\Delta \Phi = f(R,\mu)$
that has to be solved assuming that the solution vanishes at infinity. We compute the solution $\Phi$ as
\begin{equation}
\Phi(R,\mu) = -\frac{1}{2} \sum_{j=0}^N P_j(\mu) \left[ \frac{1}{R^{j+1}}E_j(R) + R^j F_j(R) \right],
\end{equation}
where
\begin{equation}
E_j(R) = \int_0^R dR^\prime {R^\prime}^{j+2} \int_{-1}^1 d \mu^\prime P_j(\mu^\prime) f (R^\prime, \mu^\prime),
\end{equation}
\begin{equation}
F_j(R) = \int_R^\infty dR^\prime \frac{1}{{R^\prime}^{j-1}} \int_{-1}^1 d\mu^\prime P_j (\mu^\prime) f(R^\prime, \mu^\prime).
\end{equation}
Equation (17) has the form
\begin{equation}
\Delta A_\phi - \frac{2 \partial_r A_\phi}{r} = g(R,\mu),
\end{equation}
which is directly related to the vectorial Poisson equation (cf.\ \cite{JMMP}). The solution $A_\phi$ that vanishes at infinity can be found as
\begin{eqnarray}
 A_\phi(R,\mu) & = & - \frac{1}{2} \sqrt{1 - \mu^2} \sum_{j=1}^N \frac{1}{j(j+1)} P_j^1(\mu) \nonumber \\
& & \left[ \frac{1}{R^j} C_j(R) + R^{j+1} D_j(R) \right],
\end{eqnarray}
where
\begin{equation}
C_j(R) = \int_0^R dR^\prime {R^\prime}^{j+1} \int_{-1}^1 \frac{d \mu^\prime}{\sqrt{1 - {\mu^\prime}^2}} P_j^1(\mu^\prime) g(R^\prime, \mu^\prime),
\end{equation}
\begin{equation}
D_j(R) = \int_R^\infty dR^\prime \frac{1}{{R^\prime}^j} \int_{-1}^1 \frac{d \mu^\prime}{\sqrt{1 - {\mu^\prime}^2}} P_j^1(\mu^\prime) g(R^\prime, \mu^\prime).
\end{equation}
Symbols $P_j$ and $P_j^1$ denote Legendre polynomials and associated Legendre polynomials of the first order, respectively. Assuming equatorial symmetry one can show that all above integrals with $P_{2j + 1}$ and $P^1_{2j}$, $j = 0, 1, 2, \dots$ vanish.

The above formulae are exact for $N \to \infty$. In our applications we set  $N \approx 100$. The integrals are computed on a spherical grid assuming a piecewise linear interpolation of $f(R,\mu)$ and $g(R,\mu)$. In most cases, the resulting integrals can be then computed analytically. Otherwise Simpson's rule is used.

The last ingredient of the numerical method is directly related to the fact that we are dealing with the free-boundary problem. In the first stage --- at the Newtonian level --- one ensures that no negative values of $h_0$ appear in the solution. At all grid points where Eq.\ (\ref{cc}) yields a negative value of $h_0$ we set $h_0 = 0$. This defines the shape of the disk at each iteration step.

A similar procedure has to be implemented at the 1PN stage. The cutoff is applied to $\rho_1$. Whenever $\rho_0 + \rho_1/c^2$ would become negative, we set $\rho_1 = -c^2 \rho_0$. Note that, in practice, corrections $\rho_1$ and $h_1$ are related by
\begin{equation}
\rho_1 = \frac{1}{K \gamma} \rho_0^{2 - \gamma} h_1.
\end{equation}
Thus $\rho_0 = 0$ implies $\rho_1 = 0$ provided that $h_1$ remains finite. 

 We use a numerical grid consisting of 200 zones in the radial direction and 800 zones in the angular one. It is important to note that disks which are relatively light in comparison to the central mass (i.e., $M_\mathrm{D} \ll m$) become slim. Consequently, it is crucial to provide sufficient angular resolution also in those cases. In a general case, our grid covers in the angular direction the region $0 \le \mu \le 1$. For thin disks we confine ourselves to $0 \le \mu \le 1/2$ or even $0 \le \mu \le 1/10$, keeping the same number of 800 angular zones. For example, for an elongated disk with $r_\mathrm{in}/r_\mathrm{out} = 1/10$ and $M_\mathrm{D}/m \approx 10^{-4}$ we get the maximal height of the disk above the symmetry plane $z = 0$, $h_\mathrm{max} \approx 0.01 r_\mathrm{out}$. In our setup this corresponds to approximately 100 angular zones occupied by the disk. This procedure was also tested for the purely Newtonian case in \cite{appb_ngc4258}, where a disk with a relative thickness of the order of $1/1000$ was computed, using up to $L = 400$ Legendre polynomials and numerical grids spanning up to $5000 \times 5000$ nodes. 

To some extent, the correctness of our numerical procedure can be tested by computing suitable virial identities. Such identities were derived in \cite{virial}. In \cite{virial} we also provided examples of convergence tests of our numerical method.

\section{Numerics:  mass ratio   and dragging of rotating systems}
\label{sec6}

\subsection{Introductory remarks}

We show in this section  that   maximal values of the total dragging and its constituents  simply scale with the 
total mass, up to  the 1PN order. 
We define normalized post-Newtonian corrections: the total  $S_\mathrm{T}\equiv |v^\phi_1/(c^2\Omega_0)|$ (see (\ref{constraint_solution_th})) correction and 
its constituent components --- the dragging (geometric) term $S_\mathrm{g} \equiv |\frac{ 3A_\phi}{4\tilde r^2\Omega_0c^2}|$, the  anti-dragging term $S_\mathrm{ad}\equiv   |\frac{3h_0}{c^2}   |$ and  the 
 centrifugal term $S_\mathrm{c}\equiv | \frac{3}{2c^2}  ( -\frac{m}{R}+ U_{0D}+\Omega_0^2\tilde r^2) |$.  It will appear  that their maximal values   almost linearly depend on  the mass functional $M_\mathrm{D}$. 
  We take care to ensure that the calculation is done well within the post-Newtonian regime: the modulus of the 1PN correction to the metric  within tori, $2|U|/c^2$, is usually smaller (or much smaller)  than 0.01 and reaches 0.05 only for heaviest disks. 
 
We assume in the numerical analysis that maximal values of  $S_\mathrm{T}, S_\mathrm{g}, S_\mathrm{ad}$ and $S_\mathrm{c}$  can occur only at the equatorial plane $z=0$. These functions satisfy elliptic equations; see (\ref{Afi}) for the equation for $A_\phi $, take into account that  $S_\mathrm{ad} \propto h_0$ and notice that $S_\mathrm{c} $ is a linear function of the enthalpy density $h_0$ (see (\ref{0Bernoulli})). The function $h_0$ satisfies equation
\begin{equation}
\Delta h_0=-4\pi \rho_0-\Omega_0^2r.
\end{equation}
We believe that by invoking the maximum (minimum) principle --- specifically, the moving planes method \cite{GWN} --- one can prove that maximal points exist at $z=0$.

Taking that  into account, and exploiting  the axial symmetry, one would need to find the maximal values of  $S_\mathrm{T}, S_\mathrm{g}, S_\mathrm{ad}$ and $S_\mathrm{c}$  in the line interval $(r_\mathrm{in},r_\mathrm{out})$ within the  equatorial section of the  disk. That is a formidable but feasible numerical  task.
  
We shall investigate disks extending from the innermost equatorial circle corresponding to  $r_\mathrm{in}=0.1, 0.4, 0.5, 0.6, 0.75$ and $0.95$  to the outermost equatorial circle  $r_\mathrm{out}=1$. The results are comprised in 10 diagrams and 2 tables --- each of graphs  demanded at least 2 dozens of numerical solutions.

It is useful to translate coordinate distances onto geometric  ones, using $R_\mathrm{S}$,  the Schwarzschild radius of the central black hole. It is assumed in all numerical calculations that the central mass is equal to 1 and the speed of light is such that $c^2=1000$. The distance areal scale is thus defined by its Schwarzschild radius $R_\mathrm{S}=2/c^2=1/500$, that  corresponds to the  isotropic coordinate radius $R=1/2000$. We constructed our toroids in limits of the 1PN order of approximation, which means that conformal factors $f$ (see the metric (\ref{sc})) are close to unity within the volumes of toroids. Thus the areal radius of the innermost equatorial circle     $r_\mathrm{in}$ is given by $R(r_\mathrm{in})=r_\mathrm{in}f^2(r_\mathrm{in})\approx r_\mathrm{in}$. Therefore inner areal radii of investigated toroids  change from  about 50 times $R_\mathrm{S}$ (for $r_\mathrm{in}=0.1$) to   $475R_\mathrm{S}$ (for $r_\mathrm{in}=0.95$).  The coordinate boundary of the whole system is located at $r_\mathrm{out}=1$  and that corresponds to the areal size of (approximately) 500 $R_\mathrm{S}$. 

We observe that the geometric contribution $\sup S_\mathrm{g}$ to the total dragging is essentially  independent of the distance of the disk from the central black hole: $\sup S_\mathrm{g}= \beta M_\mathrm{D}/m$, where the proportionality coefficient $\beta$ depends weakly only on the ratio $M_\mathrm{D}/m$ and $\beta \in (0.2, 0.5)\times 10^{-3}$.

The anti-dragging $\sup S_\mathrm{ad}$ and centrifugal $\sup S_\mathrm{c}$ terms strongly depend on the disk's distance from the black hole; they dominate at distances of the order of $50R_\mathrm{S}$ and are dwarved by the geometric dragging effect represented by $\sup S_\mathrm{g}$ at the distance of $300R_\mathrm{S}$.

\subsection{Partial  dragging effects   }

\begin{figure}[h]
\begin{center}
\includegraphics[width=0.5\textwidth]{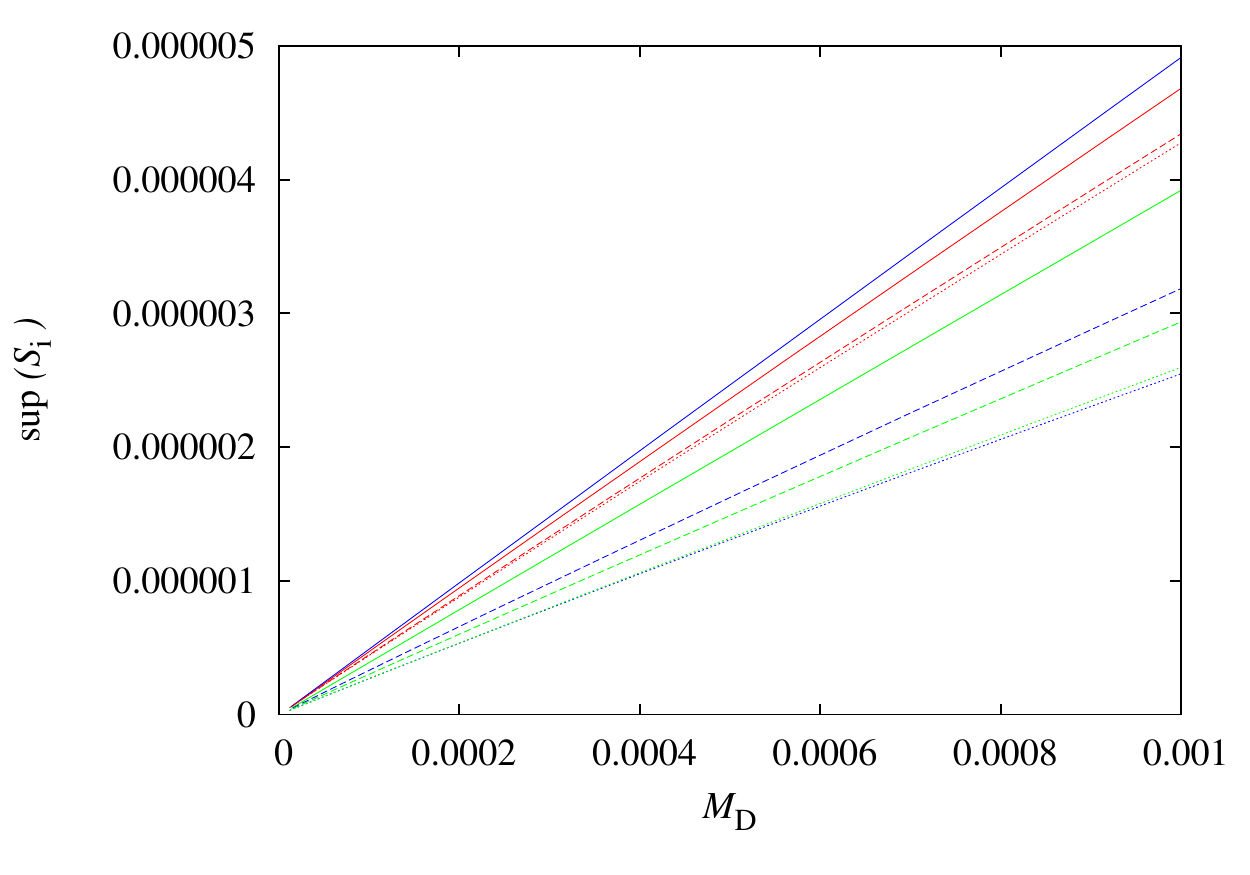}
\end{center}
\caption{\label{fig:1}
The normalized   1PN corrections   $\sup S_\mathrm{g}$ (red lines), $\sup S_\mathrm{ad}$ (blue lines) and $\sup S_\mathrm{c}$ (green lines), within the symmetry plane of the disk, put on the vertical axis --- in function of  the  mass ratio  $M_\mathrm{D}/m\le 0.001$ (displayed on the abscissa).
The inner disks's boundaries are located at $r_\mathrm{in}=0.1$ (solid lines), $ 0.4$ (broken lines)  and $0.6$ (dotted lines), respectively. The equation of state $p=K\rho^{4/3} $.}
\end{figure}

\begin{figure}[h]
\begin{center}
\includegraphics[width=0.5\textwidth]{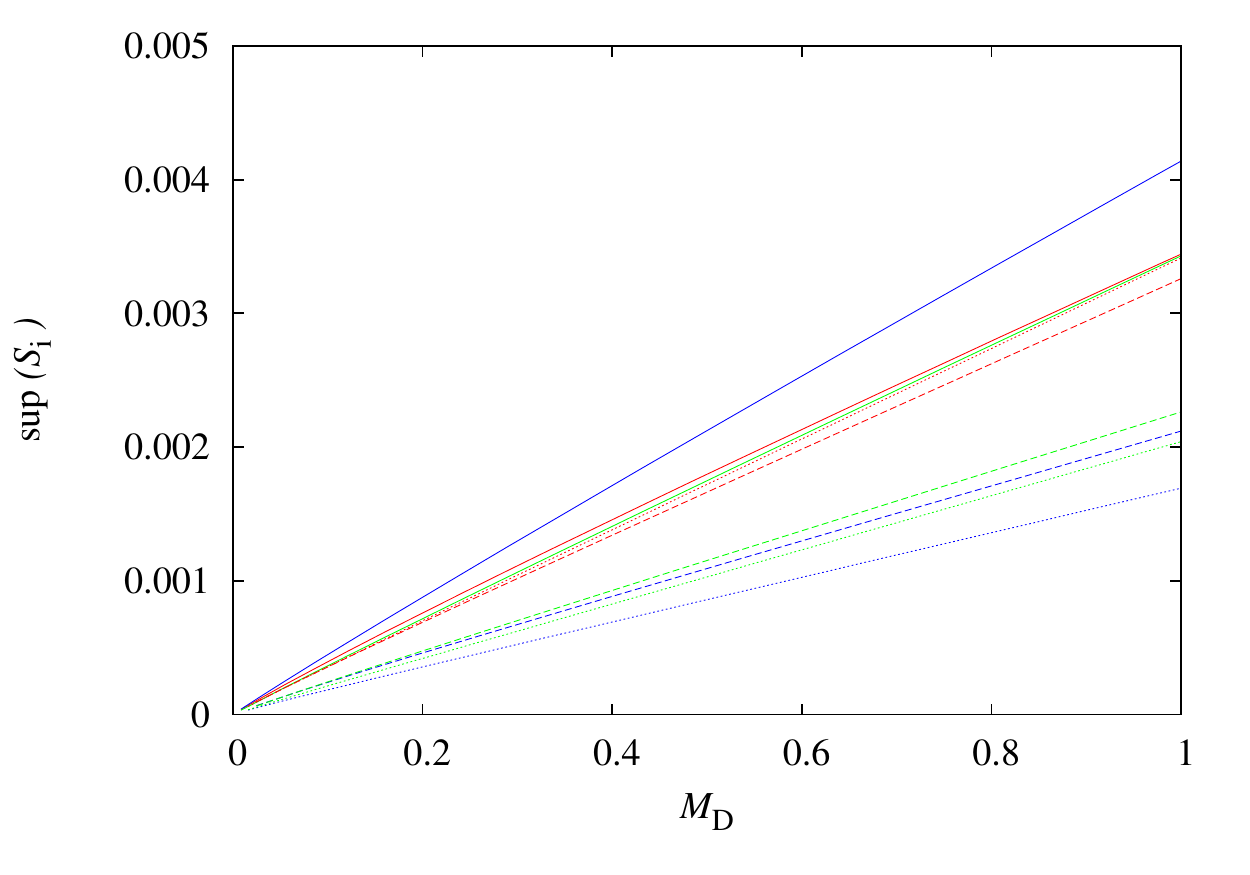}
\end{center}
\caption{\label{fig:2}
The same as in Fig. 1, but with disk masses in the interval $(0, m)$.}
\end{figure}

\begin{figure}[h]
\begin{center}
\includegraphics[width=0.5\textwidth]{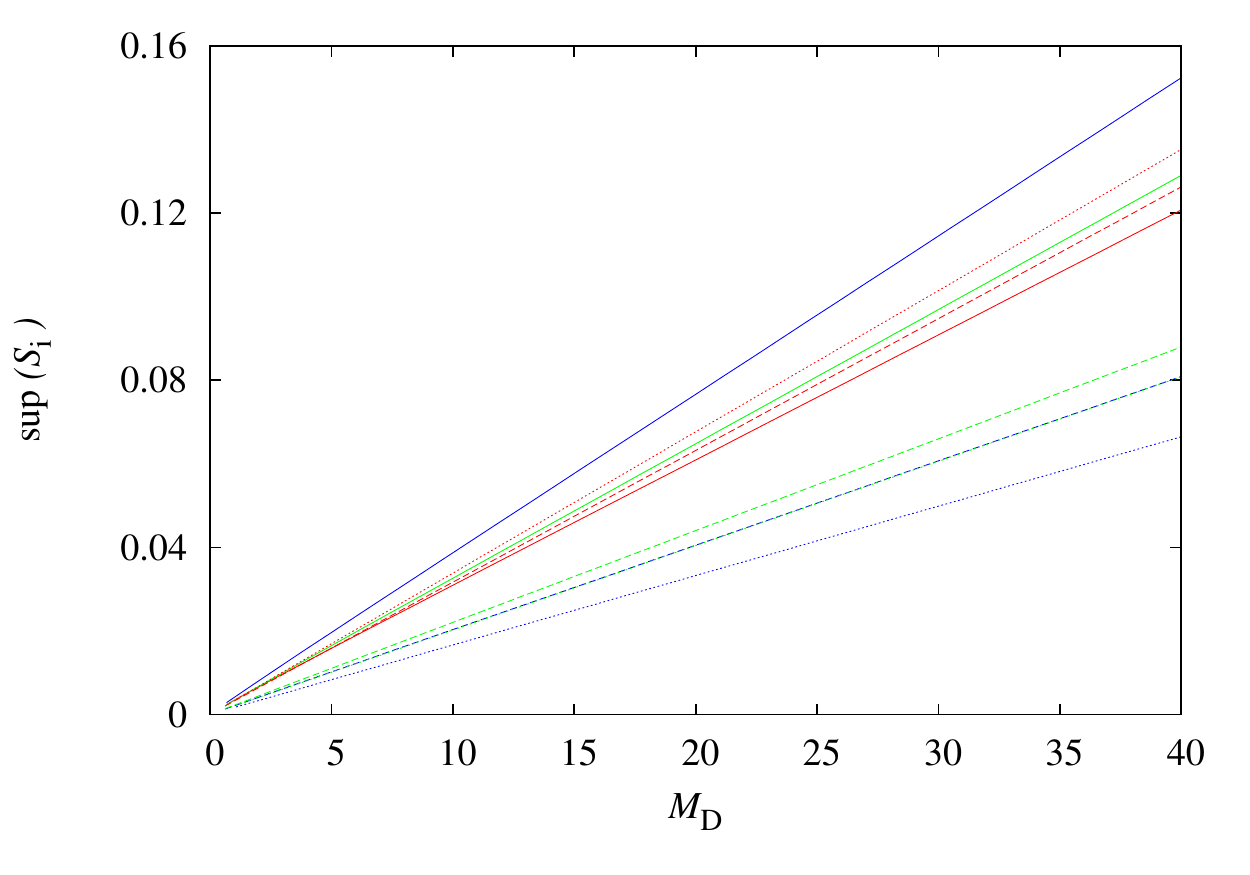}
\end{center}
\caption{\label{fig:3}
The same as in Fig. 1, but with disk masses in the interval $(m, 40 m)$.}
\end{figure}

In Figs.\ \ref{fig:1}--\ref{fig:3} we plot the dependence of the   maximal value of the 
  normalized dragging, anti-dragging and centrifugal  1PN corrections   on   the relative mass $M_\mathrm{D}/m$.  The equation of state is $p=K\rho^{4/3}$. There are a few dozens of  solutions corresponding to each of the three systems with the inner radii $r_\mathrm{in}$ situated at  $0.1, 0.4 $ and $0.6$.   Numerical data suggest  that the quantities $\sup S_\mathrm{g}$, $\sup S_\mathrm{ad}$ and $\sup S_\mathrm{c}$ are linear functions of the relative mass $M_\mathrm{D}/m$. We display the results in three diagrams. In Fig.\ 1 the ratio $M_\mathrm{D}/m$ is smaller than  $0.001$, in Fig.\ 2 we have $M_\mathrm{D}/m<1$ while in Fig.\ 3 the disk mass changes between 0 and 40 of the central mass. In each case there is an approximately linear behaviour.  

We observe that the slope coefficient of the geometric dragging quantity $\sup S_\mathrm{g}/M_\mathrm{D}$ is only weakly  dependent on the position  of a disk ---  it   changes   slightly  with the change of the  inner radius $r_\mathrm{in}$. In contrast to that, the slopes of the anti-dragging and centrifugal objects $\sup S_\mathrm{ad}/M_\mathrm{D}$ and $\sup S_\mathrm{c}/M_\mathrm{D}$  are quite  sensitive to the disk location; they  become significantly smaller with the increasing  distance from the central black hole. For the case $r_\mathrm{in}=0.1$ we have $\sup S_\mathrm{ad} > \sup S_\mathrm{c} > \sup S_\mathrm{g}$; clearly the anti-dragging effect dominates, for all masses within the range $M_\mathrm{D}\in (10^{-4}m, 40m)$. With the increase of the distance the
picture  reverses;   at $r_\mathrm{in}=0.6$ we have $\sup S_\mathrm{g} > \sup S_\mathrm{c} > \sup S_\mathrm{ad}$.  One can notice, choosing two disks of the same mass but with different inner boundaries,  that the value of $\sup S_\mathrm{ad}$ at $r_\mathrm{in}=0.6$  is about one half of its value at  $r_\mathrm{in}=0.1$.

One can read from these diagrams, that for the equation of state $p=K\rho^{4/3} $ 
and $M_\mathrm{D} <10^{-3}m$, the correction  $\sup S_\mathrm{g}\approx  4\times 10^{-3} M_\mathrm{D}/m$ (see Fig.\ 1); the same  equation of state corresponds to  $\sup S_\mathrm{g}\approx  ({\textcolor{red}{2.8-3.2}})\times 10^{-3} M_\mathrm{D}/m$, where $M_\mathrm{D} \in (m, 40 m) $ (see Fig.\ 3).

\begin{figure}[h]
\begin{center}
\includegraphics[width=0.5\textwidth]{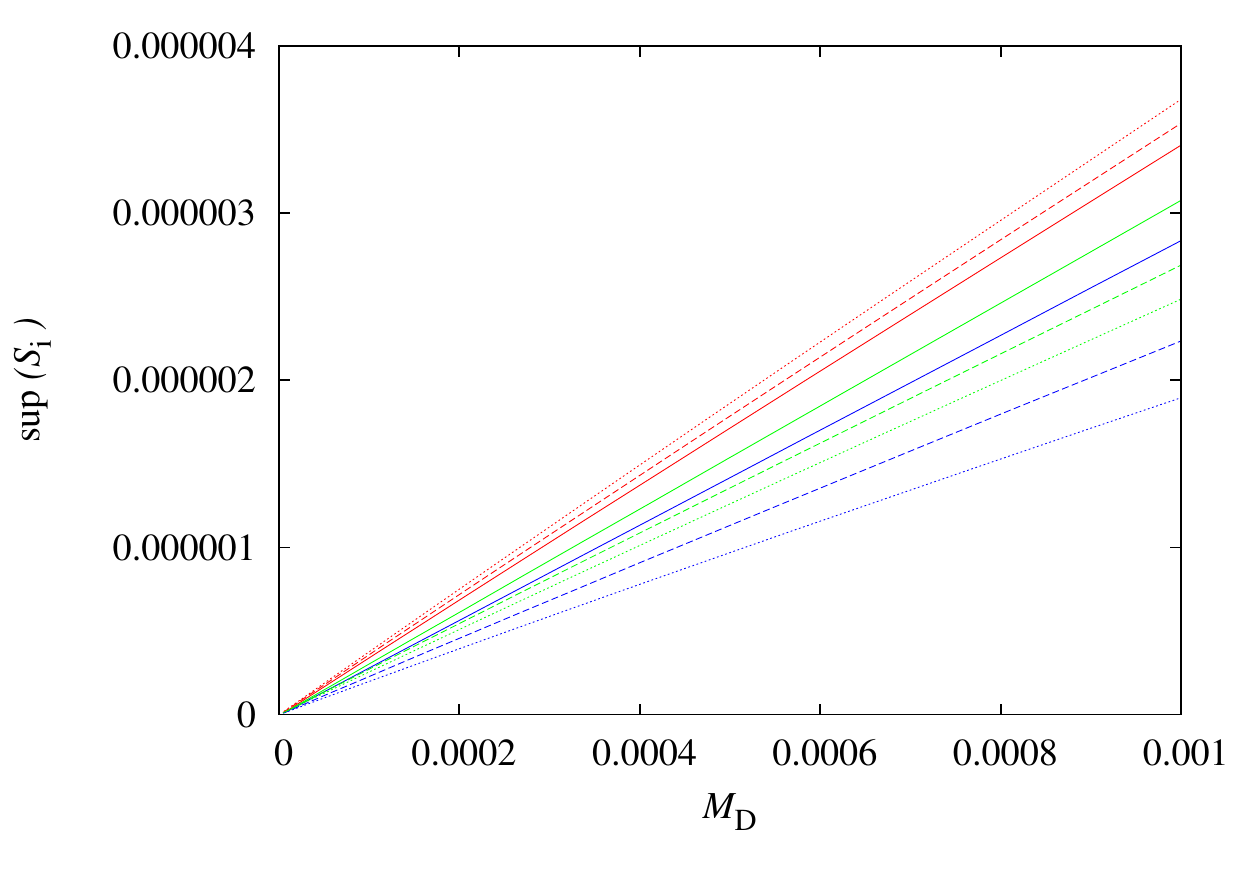}
\end{center}
\caption{\label{fig:4} The normalized 1PN corrections $\sup S_\mathrm{g}$ (red lines), $\sup S_\mathrm{ad}$ (blue lines) and $\sup S_\mathrm{c}$ (green lines), within the symmetry plane of the disk, put on the ordinate ---   
 in function of  the  mass ratio  $M_\mathrm{D}/m\le 0.001$ (displayed on the abscissa).
The inner disks's boundaries located at $r_\mathrm{in}=0.1$ (solid lines), $ 0.4$ (broken lines)  and $0.6$ (dotted lines), respectively. The equation of state is $p=K\rho^{5/3}$.}
\end{figure}
 
\begin{figure}[h]
\begin{center}
\includegraphics[width=0.5\textwidth]{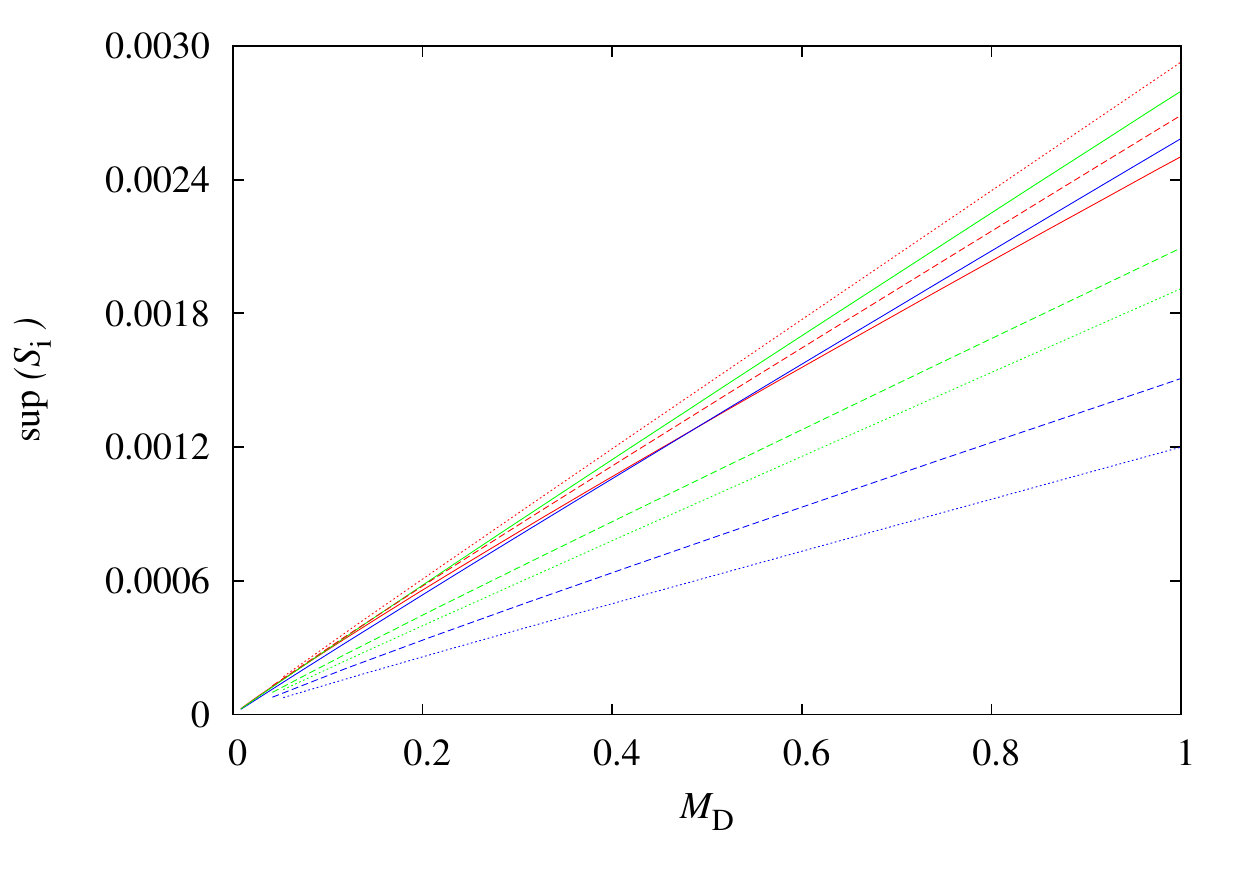}
\end{center}
\caption{\label{fig:5}The same as in Fig. 4, but with   disk masses in the interval $(0, m)$.}
\end{figure}

\begin{figure}[h]
\begin{center}
\includegraphics[width=0.5\textwidth]{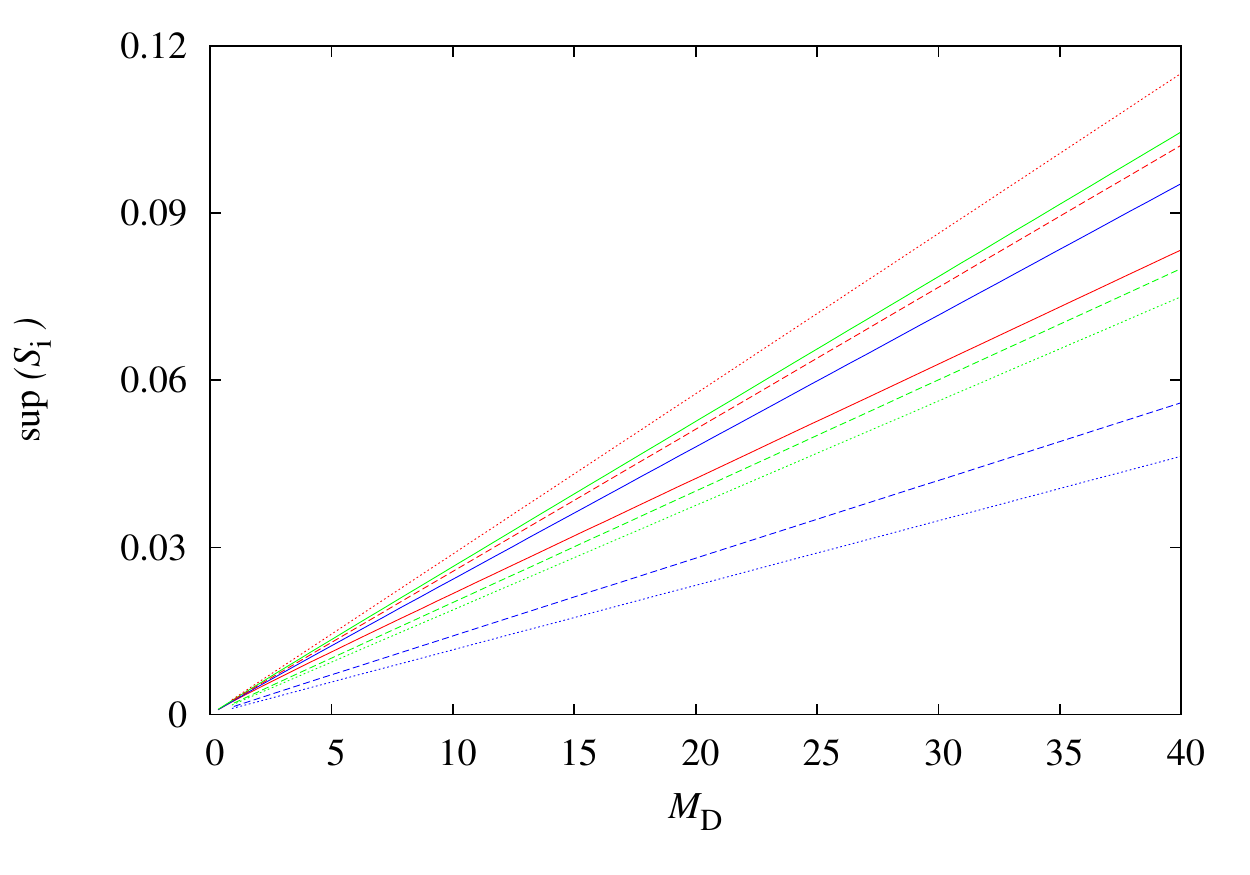}
\end{center}
\caption{\label{fig:6}
The same as in Fig. 4, but with   disk masses in the interval $(m, 40m)$.
 }
\end{figure}

This analysis yields similar results also for disks with the polytropic equation of state  $p=K\rho^{5/3}$  (see Figs.\ \ref{fig:4}--\ref{fig:6}), with  several notable differences.
In the mass interval $M\in (0, 10^{-3})m$ and $r_{in}=0.1$ we observe $S_\mathrm{g}>S_\mathrm{c} > S_\mathrm{ad}$, while the  centrifugal term $\sup S_\mathrm{c}$ dominates at $r_\mathrm{in} = 0.1$ for higher   disk masses.  The anti-dragging related object $\sup S_\mathrm{ad}$ exceeds the geometric part $\sup S_\mathrm{g}$ only for $M_\mathrm{D}>0.4m$, again only at $r_\mathrm{in}=0.1$. When $r_\mathrm{in}=0.4$ or $0.6$, then always geometric effects dominate over centrifugal ones, and those anti-dragging ones are the weakest of all.  
 
\subsection{Total dragging: a linear function of relative  mass}

\begin{figure}[h]
\begin{center}
\includegraphics[width=0.5\textwidth]{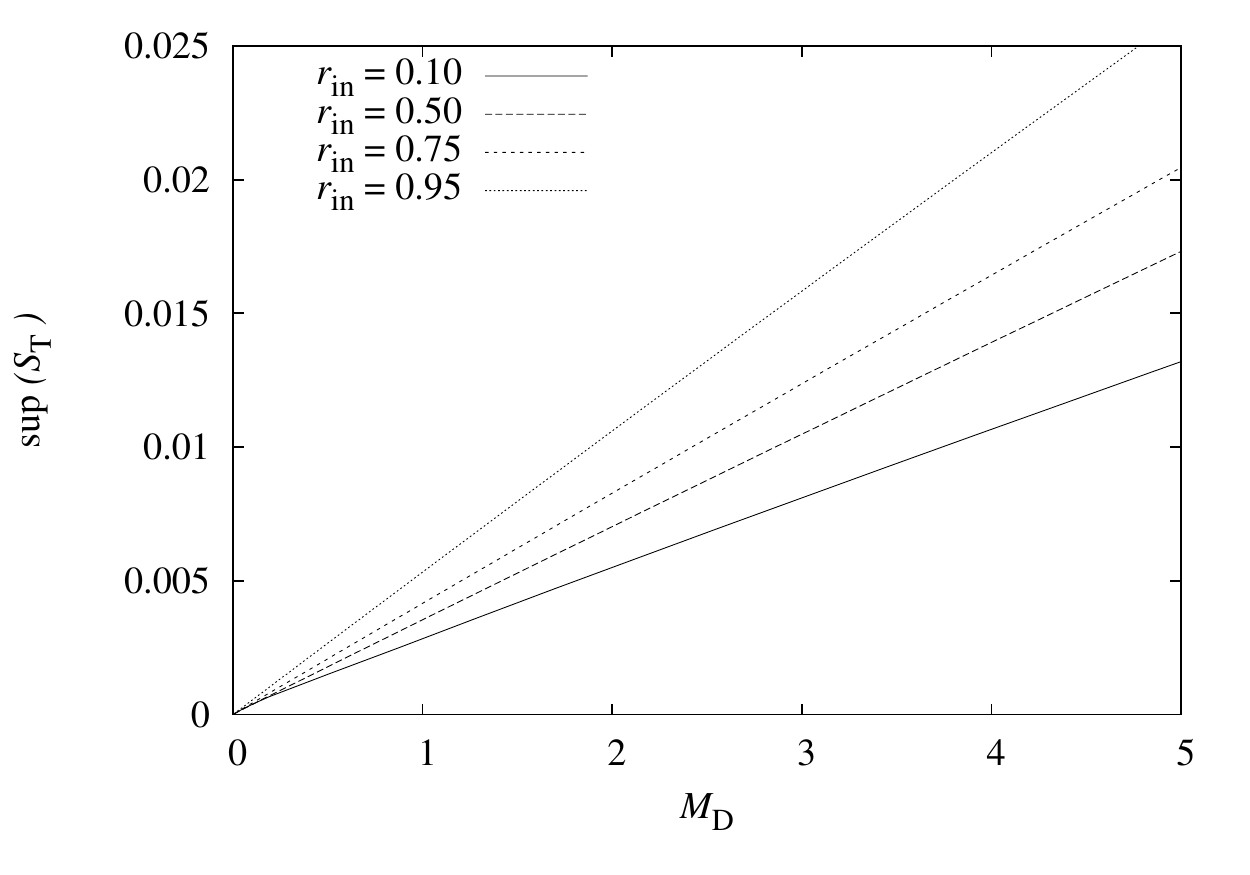}
\end{center}
\caption{\label{fig:7}
The maximum of the     normalized 1PN correction   $\frac{v_1^\phi}{\Omega_0c^2} $ within the disk (on the ordinate)   in function of  the asymptotic mass of the disk $M_\mathrm{D}$ (plotted on 
the abscissa),
for disks with inner boundaries located at $r_\mathrm{in}=0.1, 0.5, 0.75$ and $0.95$, respectively.
The equation of state is  $p=K\rho^{4/3}$.
 }
\end{figure}

\begin{figure}[h]
\begin{center}
\includegraphics[width=0.5\textwidth]{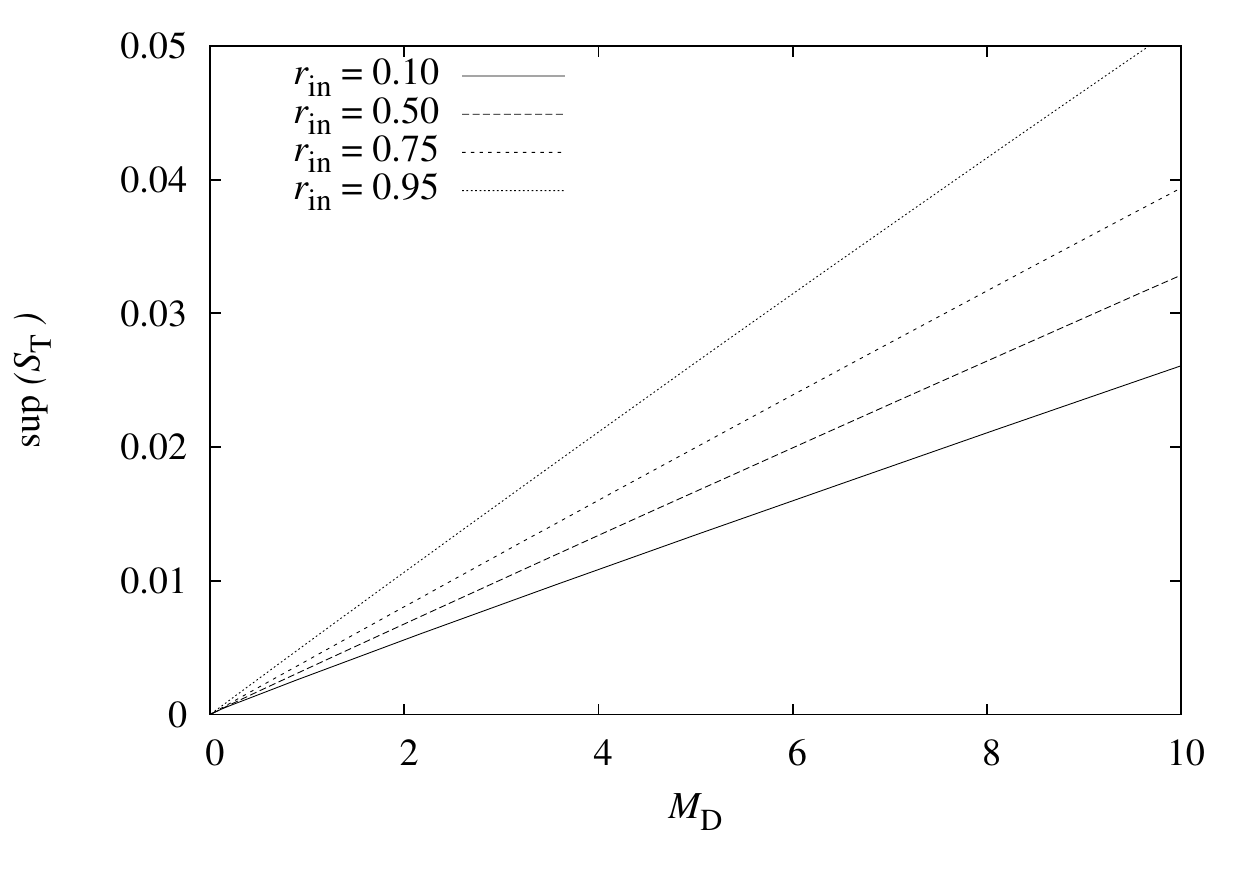}
\end{center}
\caption{\label{fig:8}
The same as in Fig. 7, but for the equation of state $p=K\rho^{5/3}$.
 }
\end{figure}

In  Figs.\ \ref{fig:7} and \ref{fig:8} we plot the dependence of the   maximal value of the 
  normalized 1PN correction   $S_\mathrm{T}=\frac{v_1^\phi}{\Omega_0c^2} $ on the relative mass $M_\mathrm{D}/m$.   

 We have found several dozens of solutions corresponding to each of the four systems with the inner radii $r_\mathrm{in}$ situated at 
$0.1, 0.5, 0.75$ and $0.95$. Again it appears  that $\sup S_\mathrm{T}$ is a strictly linear function of the relative mass.  For the sake of clarity, we 
show corresponding results in two diagrams. In Fig.\ 7, for the equation of state $p=K\rho^{4/3}$,  the mass of the disk is smaller than $5m$. The most interesting feature of these plots is that the steepest line correspond to the farthest disk; for a fixed mass of disks the maximal value of the normalized correction   $S_\mathrm{T}  $ increases with the increase of $r_\mathrm{in}$. 
  
In Fig.\ 8, which corresponds to  the equation of state $p=K\rho^{5/3}$,  the disk mass changes between 0 and 20 $m$. The steepness of the lines is almost the same as in the Fig.\ 7, and again the steepest lines correspond to the farthest disk.

A closer inspection in the regime  of light disks, with masses significantly smaller than the central mass,  reveals a more complex picture (see Fig.\ 9). For light disks  --- our rough estimate is $M_\mathrm{D}<0.2m$ ---    the normalized dragging $\sup S_\mathrm{T}$  for disks with inner boundaries located at $r_\mathrm{in}$ may decrease with the increase of $r_\mathrm{in}$, and then  it can start to increase. The value of the critical radius $r_\mathrm{cr}$, where the behaviour changes, depends on the mass $M_\mathrm{D}$, but is probably  smaller than half of the central mass. In the situation displayed on Fig.\ 9, we observe that      $\sup S_\mathrm{T}(r_\mathrm{in}=0.1)$ (solid line) intersects  with $\sup S_\mathrm{T}(r_\mathrm{in}=0.5)$ (long broken line), at masses (roughly) 0.01 $m$ and 0.12 $m$. That means that disks with masses in the interval $(0.01 m, 0.12 m)$ are characterized by functions $\sup S_\mathrm{T}$  that are not monotonic as  functions of $r_\mathrm{in}$, in the region $0.1<r_\mathrm{in} <0.5$.

\begin{figure}[h]
\begin{center}
\includegraphics[width=0.5\textwidth]{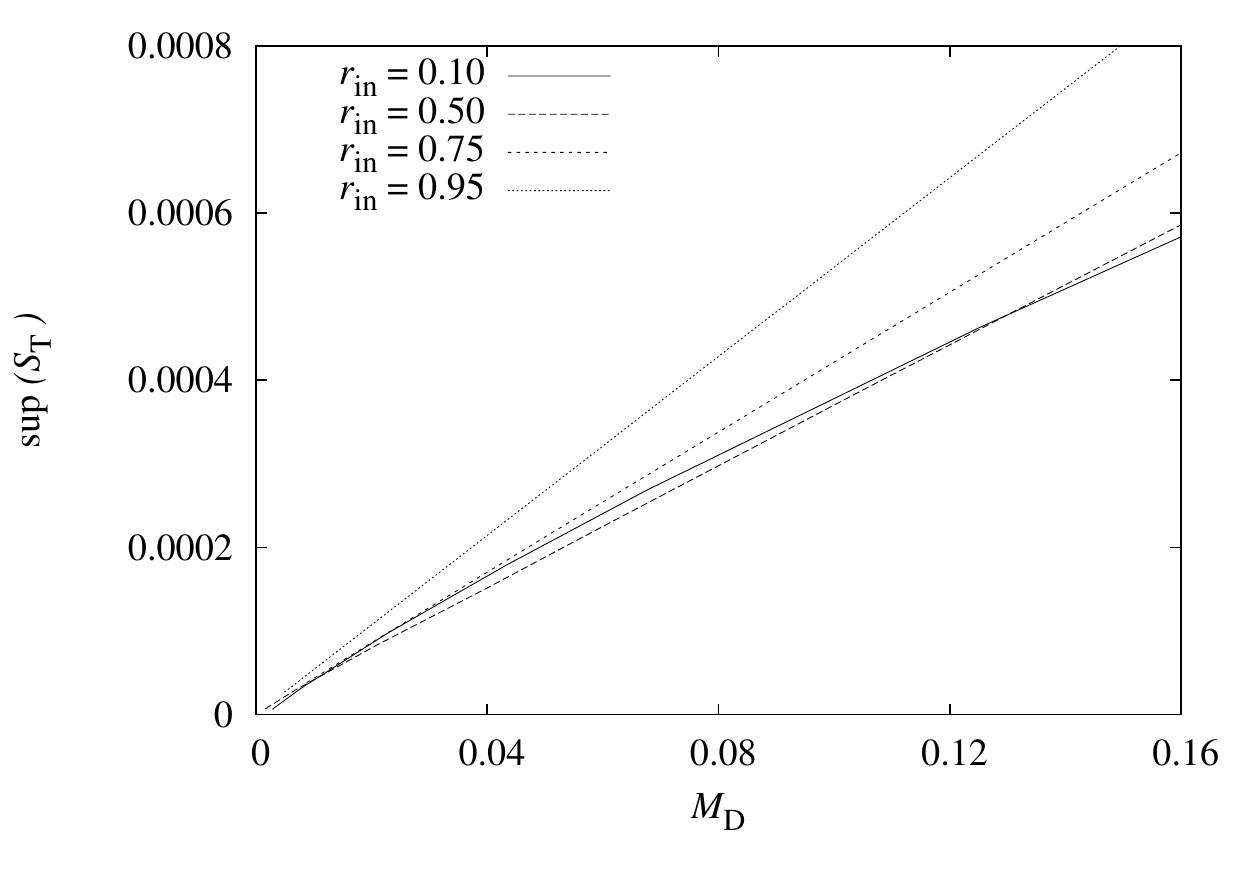}
\end{center}
\caption{\label{fig:9}
The values of $\sup S_\mathrm{T}$ are put on  ordinate, while disk masses  are put on the abscissa. The two lines 
corresponding to $r_\mathrm{in}=0.1$  (solid line) and $r_\mathrm{in}=0.5$ (broken line) cross  around $M_\mathrm{D}=0.11m$. The equation of state is $K\rho^{4/3}$.  }
\end{figure}

This behaviour is due to the rapid growth  of the anti-dragging  term $\sup S_\mathrm{ad}$  with the increase of mass that compensates a moderate growth of the dragging quantity $\sup S_\mathrm{g}$ (compare  Figs.\  1 and 2), for   disk masses that are small enough.   That would cause the falloff of the normalized correction $\sup S_\mathrm{T}$ in an interval
of small radii $r_\mathrm{in}$.

\section{Angular momentum: local   versus total  }
\label{sec7}

In this section we address issues concerning the dragging at the black hole horizon and the  distribution     of angular momentum in systems with rotating rings. Let us recall that Nishida and Eriguchi \cite{nishida_eriguchi} applied the rotation law $j(\Omega )=A(\Omega  -B)$. They found, in particular,  that the angular momentum of  the central  black hole can vanish and --- more generally --- its internal spinning  parameter  $a_\mathrm{S} $ can be both negative  and positive. For some configurations one would have $|a_\mathrm{S}|>1$ and the dragging function $A_\phi $ would  vanish at the horizon. Ansorg and Petroff \cite{Ansorg_Petroff} assumed the constant angular velocity within the disk and they also obtained central black holes   with $|a_\mathrm{S}|>1$.

Our results are comprised in Table 1. The internal spinning parameter $a_\mathrm{S} $ is contained within the range $(10^{-4},10^{-10})$. That means that  the central black hole  can be safely approximated by  a Schwarzschild black hole. We already proved that the metric function $A_\phi $ can have only isolated zeroes \cite{JMMP}; that is, it cannot vanish at the horizon of the black hole. There is an apparent  discrepancy between our results and those of  \cite{nishida_eriguchi,  Ansorg_Petroff, shibata}.  This can be ascribed  mainly   to the fact, that our central black hole is inherently spinless, in contrast to  what was assumed in quoted papers, in which central black holes possessed their own internal spin. We think, however,    that a different  choice of   the rotation law --- the uniform rotation  \cite{Ansorg_Petroff}, the constant $j$ adopted by Shibata \cite{shibata} and the linear law $j(\Omega )=A(\Omega  -B)$ \cite{nishida_eriguchi}, within the fully general-relativistic equilibrium  ---  instead of the  Keplerian rotation law (\ref{momentum}) in the perturbation equilibrium  (up to the 1PN order) --- also played a role. It might happen that for the post-Newtonian descriptions with  $\Omega = \mathrm{const}$ or $j = \mathrm{const}$  the apparent horizon's spinning parameter would   be much larger than observed here.  We cannot exclude also that  the fully general-relativistic equilibria with the Keplerian law (\ref{momentum}) would have  values of  $a_\mathrm{S} $ exceeding  1.  

Figure 10 shows the dimensionless spinning parameter for the whole spacetime,  $a_\infty \equiv  \frac{cj_\infty }{M^2} $, where  $M$ is the total mass  read off from the asymptotic behaviour of the total potential $\frac{-m}{R}+U_\mathrm{D}$.  Let us remark that for the Schwarzschild black hole  $a_\infty =0$, while for the extremal Kerr black hole $a_\infty =1$. All nonextremal Kerr black holes have $0<a_\infty <1$. In our case we have values of $a_\infty $ that are large for disks heavier than the black hole   and that are small in the opposite case, when $M_\mathrm{D}\ll m$.   Our numerical data clearly demonstrate that the   spinning parameter  in non-Kerr  stationary configurations can be significantly larger than 1.

The  most significant  observation is that only a tiny fraction $j_\mathrm{S}$ of the angular momentum is deposited within the central black hole. The ratio $j_\mathrm{S}/j_\infty $ varies from $10^{-6}$ for the disk closest to the center, with the equatorial inner edge located at $r_\mathrm{in}=0.1$ (which corresponds to about 50 $R_\mathrm{S}$) to $10^{-12}$ for systems with  the equatorial inner edge placed at  $r_\mathrm{in}=0.95$ (that corresponds to  about 475 $R_\mathrm{S}$).

\begin{figure}[h]
\begin{center}
\includegraphics[width=0.5\textwidth]{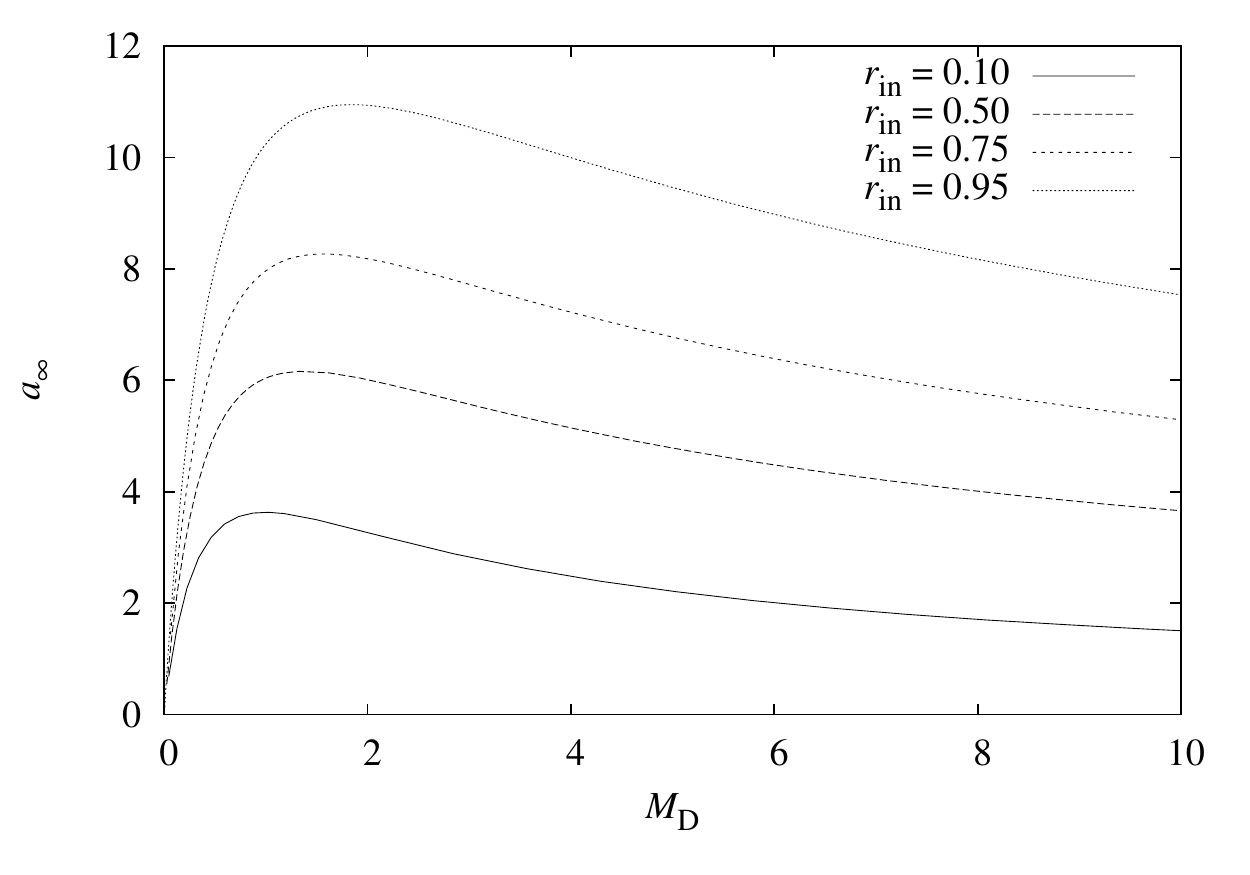}
\end{center}
\caption{\label{fig:10}
Asymptotic values   of  the spin parameter $a_\infty \equiv \frac{cj_\infty }{M^2} $, where $M=m+M_\mathrm{D}$ is the asymptotic mass.}
\end{figure}

\begin{table}
\caption{\label{Table1}The asymptotic angular momentum (the first column) and the black hole angular momentum for disks with $r_\mathrm{in}=0.1$ (the third column). Fourth column gives the disks's mass in units of the central mass $m$.  The second column displays   the area of the central black hole   in terms of $R_\mathrm{S}\equiv 2m/c^2$. The last column is the maximal height of the disk.} 
\begin{ruledtabular}
\begin{tabular}{c c c c c}
$L$ & $A_\mathrm{AH}/16 \pi$ & $j_\mathrm{S}$ & $M_\mathrm{D}$ & $h_\mathrm{max}$ \\
0.6880 & 0.9972 & 6.034 $\times 10^{-10}$ & 1.493 & 0.3922 \\
 1.006 & 0.9959 & 1.339 $\times 10^{-10} $ & 2.156 & 0.3922 \\
 1.351 & 0.9945 & 2.459 $\times 10^{-9}   $ & 2.848 & 0.4269 \\
 1.723 & 0.9931 & 4.018 $\times 10^{-9}   $ & 3.561 & 0.4529 \\
 2.119 & 0.9917 & 6.065 $\times 10^{-9}   $ & 4.290 & 0.4728 \\
 2.539 & 0.9903 & 8.644 $\times 10^{-9}   $ & 5.032 & 0.4883 \\ 
 2.981 & 0.9888 & 1.180 $\times 10^{-8}   $ & 5.783 & 0.5012 \\
 3.445 & 0.9873 & 1.557 $\times 10^{-8}   $ & 6.544 & 0.5120 \\
 6.071 & 0.9798 & 4.543 $\times 10^{-8}   $ & 10.43 & 0.5211 \\
 9.174 & 0.9722 & 9.539 $\times 10^{-8}   $ & 14.42 & 0.5515 \\
 12.72 & 0.9645 & 1.709 $\times 10^{-7}   $ & 18.50 & 0.5688 \\
 16.70 & 0.9568 & 2.755 $\times 10^{-7}   $ & 22.64 & 0.5799 \\
 21.09 & 0.9490 & 4.134 $\times 10^{-7}   $ & 26.84 & 0.5878 \\
 25.89 & 0.9413 & 5.886 $\times 10^{-7}   $ & 31.11 & 0.5935 \\
 31.08 & 0.9335 & 8.051 $\times 10^{-7}   $ & 35.43 & 0.6014 \\
 36.67 & 0.9258 & 1.067 $\times 10^{-6}   $ & 39.82 & 0.6044 \\
 \end{tabular}
 \end{ruledtabular}
\end{table}

\section{Angular momentum and isoperimetric inequalities}
\label{sec8}

We shall start with a  compendium on  various concepts of  mass and mass densities.
In formula (\ref{emD}) appears the baryonic mass density $\rho $. It plays the role of an 
integration factor, ensuring the conservation of the baryonic current $\rho u^\mu $: $\nabla_\mu (\rho u^\mu )=0$.  
The volume integral $\int_VdV \rho $ is the baryonic mass --- a quantity that is conserved.
In the Newtonian limit the baryonic mass coincides with asymptotic mass, but within General Relativity the asymptotic (Bondi-Einstein-Landau-Lifschitz-Freund-Trautman-Arnowit-Deser-Misner) mass (see \cite{CJMC} for a discussion)   is distinct from the baryonic mass.  In our foregoing considerations we always dealt with the conserved (ADM) asymptotic mass. One defines also
the rest energy density $e \equiv  \rho (c^2+h) -p= \rho c^2 +\frac{p}{\gamma -1} $, 
the total rest energy    $E_\mathrm{r}\equiv \int_VdVe$ \cite{KMRS} and  the related  rest mass $M_\mathrm{r}(V) \equiv \frac{1}{c^2} \int_VdVe $ (see, for instance, \cite{KMRS}). 

One defines the momentum density   $J_\nu \equiv t_\mu T^\mu_\nu $.
It is well known that perfect fluids with the  polytropic  exponent $\gamma \le 2$ satisfy the dominant energy condition $e \geq \sqrt{J^k J_k}\equiv |J|$
\cite{Wald}, that reduces in our case to the inequality
\begin{equation}
\label{DEC}
e \geq \sqrt{J^\phi J_\phi}  .
\end{equation}

   It was already shown that --- assuming the dominant energy condition,  conformal flatness and a kind of convexity ---   the total rest mass $M_\mathrm{r}(V)$  can be bounded by  $2c^2l(V)$, where $l$ is a geodesic size of the configuration \cite{Malec1991, M91, KM, BMO90, BM89, MX}. These are special   cases, but they  are in a sense more general than it is needed for our purpose, since their derivation does not  require the assumption of stationarity, adopted in this paper; it is enough to guarantee that the configurations are momentarily static. On the other hand, one needs a regular center and a kind of convexity, that can be obeyed by rotating stars,  but which is not valid for rotating toroids. Thus  application of  estimates of the type    
\begin{equation}
M_\mathrm{r}(V)\le 2c^2l(V) 
\label{iso}
\end{equation}
--- bounding  the rest mass in terms of geodesic radii ---  to  toroidal  systems  would require a renewed analysis.

Dain recently investigated  two other size measures, related to the so-called radius $R_\mathrm{SY}$ of Schoen and Yau \cite{SY}. One of them is  defined as 
\begin{equation}
R^\prime \equiv \frac{2\sqrt{\int_V|\eta |dV}}{\pi R_\mathrm{SY}}
\label{1dain}
\end{equation}
while the other is given by
\begin{equation}
\hat R \equiv \frac{2\sqrt{\int_V|\eta |dV}}{\pi R_\mathrm{OM}};
\label{2dain}
\end{equation}
 here $R_\mathrm{OM}$ is  a  modification of the $R_\mathrm{SY}$ measure due to  N.\ \'{O} Murchadha   \cite{NOM}.
 They are formulated in terms of quantities relating entirely to toroids; they do not assume convexity and they can be applied to systems investigated in this paper.
   
The total angular momentum can be written as  
$L= c^{-1} \int_V J_\nu \eta^\nu  dV = c^{-1} \int_V J_\phi   dV$.   
 The  application of (\ref{DEC}) to  the definition of the angular momentum (\ref{volume_ang})  gives a string of inequalities
  \begin{eqnarray}
|L| & \leq & \frac{1}{c} \int_V |J| |\eta| dV \leq \frac{1}{c} \sup_V|\eta|\int_V |J| dV \nonumber \\
& \leq & c\sup_V|\eta| M_\mathrm{r}(V).
\label{Jestimate}
\end{eqnarray}
In the last inequality, we have used the assumption that the data satisfy the dominant energy condition.  Provided that $M_\mathrm{r}(V)\le 2c^2l(V) $ and   that we are in the perturbative regime (which means that the conformal factor is close to 1, we get $\sup_V|\eta| \approx R(V)$, where $R(V)=C/(2\pi )$ is the areal size of a toroid; its circumference $C$ divided by $2\pi $.
This leads to the inequality, that is valid for rotating toroids, supposing conformal flatness and the perturbative regime, 
 \begin{equation}
\label{Jestimate1}
|L_z| \lesssim   2c^3 R(V) l(V).
\end{equation}
 
This derivation of  (\ref{Jestimate1}) is analogous to  that of  S. Dain \cite{Dain, Daingrg} for axially symmetric systems, without postulating stationarity, but assuming an isoperimetric inequality as in (\ref{iso}).

 Dain has got   another  bound onto a local angular momentum  within a finite volume,  that does not require 
 postulating any isoperimetric inequalities but instead assumes   constant density bodies, of the form 
 \begin{equation}
\tilde R^2\ge \delta \frac{1}{c^3}|L|
\label{Dain}
\end{equation}
Here $\delta =\frac{24}{\pi^3}$ is a coefficient of the order of unity and $\tilde R =\hat R$ or $\tilde R =R^\prime $.
We have to note that, unfortunately, rotating disks are not characterized by constant mass densities.

M. Khuri obtained a similar upper bound, dropping the assumption of constant density but imposing a stronger energy condition and a strong un-trapped condition \cite{MKh}.  

Tables I and II show results of our numerical calculations.  Columns 1 and 3  show values of angular momentum of the whole system and of the black hole, respectively; clearly, the angular momentum deposits in peripheral regions. Column 2 shows   values of the control parameter $c_\mathrm{p}\equiv \frac{\int_Sd^2S}{4\pi R^2_\mathrm{S}}$; all its entries should be close to 1, since that means that the horizon is indeed located at the coordinate radius $R\approx \frac{m}{2c^2}$, as assumed in the numerical calculation. Column 4 shows the mass of the disk in terms of the central  mass $m$. The last column presents the coordinate height  of the disk. Let us remind that we assumed $c^2=1000$.   

The validity of (\ref{Jestimate1}) is    expected, since it can be proven in the 1PN order of approximation, but the fact that it is satisfied with a huge margin may be interpreted as suggesting the universality  of the isoperimetric inequality (\ref{iso}). 

The geometry inside toroids is approximately Euclidean, hence the Schoen and Yau radius  $R_\mathrm{SY}$ is roughly equal to one half of the $\min \left( \frac{1}{2} (r_\mathrm{out}-r_\mathrm{in}), h\right) $; it is easy to check that  $R_\mathrm{SY}^2\gg \frac{1}{c^3}|L|$ for all systems that are described in Table I, while  $R_\mathrm{SY}^2\ll \frac{1}{c^3}|L|$ for configurations listed in Table II. Both measures $R^\prime $ and $\hat R$ are much greater than $R_\mathrm{SY}$ (they are of the order of $R(V)$, or of the radius of the great circle of the toroid) and the inequality (\ref{Dain}) holds true for both of them, and for all systems displayed in the two Tables. That might be regarded as surprising, since --- as we pointed above  ---  stationary selfgravitating toroids do not satisfy the basic condition  of \cite{Dain}, that the mass density is constant; that probably means that a better analytic estimate should be available under much weaker  suppositions.   

\begin{table}
\caption{\label{Table2}The asymptotic angular momentum (the first column) and the black hole angular momentum for disks with $r_\mathrm{in}=0.95$ (the third column). Fourth column gives the disks's mass in units of the central mass $m$.  The second column  depicts the area of the central black hole radius  in terms of $R_\mathrm{S}\equiv 2m/c^2$. The last column is the maximal height of the disk.   } 
\begin{ruledtabular}
\begin{tabular}{c c c c c}
$L$ & $A_\mathrm{AH}/16 \pi$ & $j_\mathrm{S}$ & $M_\mathrm{D}$ & $h_\mathrm{max}$ \\
1.530 & 0.9988 & 4.176 $\times 10^{-11}  $ & 1.145 & 0.033391 \\
2.563 & 0.9983 & 1.048 $\times 10^{-10}  $ & 1.722 & 0.033395 \\
3.775 & 0.9977 & 2.037 $\times 10^{-10}  $ & 2.302 & 0.033395 \\
5.061 & 0.9971 & 3.435 $\times 10^{-10}  $ & 2.885 & 0.033395 \\
6.503 & 0.9965 & 5.287 $\times 10^{-10}  $ & 3.470 & 0.033395 \\
8.062 & 0.9959 & 7.632 $\times 10^{-10}  $ & 4.059 & 0.033395 \\ 
9.733 & 0.9953 & 1.051 $\times 10^{-9}    $ & 4.650 & 0.033395 \\ 
11.15 & 0.9947 & 1.396 $\times 10^{-9}    $ & 5.243 & 0.033395 \\ 
13.39 & 0.9942 & 1.800 $\times 10^{-9}    $ & 5.839 & 0.033395 \\ 
15.37 & 0.9936 & 2.269 $\times 10^{-9}    $ & 6.438 & 0.033395 \\ 
17.45 & 0.9930 & 2.804 $\times 10^{-9}    $ & 7.040 & 0.033395 \\ 
19.62 & 0.9924 & 3.409 $\times 10^{-9}    $ & 7.645 & 0.033395 \\ 
21.88 & 0.9918 & 4.086 $\times 10^{-9}    $ & 8.252 & 0.033395 \\ 
24.23 & 0.9912 & 4.839 $\times 10^{-9}    $ & 8.862 & 0.033395 \\ 
26.67 & 0.9906 & 5.670 $\times 10^{-9}    $ & 9.474 & 0.033395 \\ 
29.20 & 0.9901 & 6.583 $\times 10^{-9}    $ & 10.09 & 0.033395 \\
31.81 & 0.9895 & 7.579 $\times 10^{-9}    $ & 10.70 & 0.033395 \\ 
34.50 & 0.9889 & 8.661 $\times 10^{-9}    $ & 11.33 & 0.033395 \\ 
37.28 & 0.9883 & 9.832 $\times 10^{-9}    $ & 11.95 & 0.033395 \\ 
 \end{tabular}
 \end{ruledtabular}
\end{table}

\section{Concluding remarks}
\label{sec9}

We have demonstrated, that in Keplerian systems consisting of a rotating toroid and a  {\it spinless} black hole, the  black hole can be (essentially) Schwarzschildean --- almost all angular momentum is deposited within the toroid. This is true for a large spectrum of systems, for disk masses $M_\mathrm{D}\in (10^{-4}m, 40m)$ ($m$ is the black hole mass).  That  observation would mean, that there  is a need  to do a more careful interpretation of  those  astrophysical objects with black holes where the Keplerian rotation curve is observed.  The standard  practise is to assume that the black hole is Kerr-like and that the toroid is test-like, that is its self-gravity can be neglected.   Our results suggest that an alternative picture is plausible, with the central black hole being Schwarzschildean and the   disk carrying angular momentum and selfgravitating, even for light disks, $M_\mathrm{D}/m \ll 1 $.

 Our numerics suggests that there is a need to include the all \textit{three} weak field components 
of the general-relativistic effects \cite{MM2015}  in gaseous disks circulating around a spinless black hole. The geometric (frame-dragging) effect becomes dominating at relatively large distances; the two other effects can contribute up to 50\% even at distances $R\approx 500 R_\mathrm{S}$, and even for light disks, $M_\mathrm{D}\ll m$.  

One of the main surprises in this investigation is the fact, that all weak general-relativistic components  (dragging, anti-dragging and the centrifugal) scale with $M_\mathrm{D}/m$; their maximal values are  proportional to $M_\mathrm{D}/m$. The same is true concerning the total 1PN correction to the angular velocity.  We do not know any simple explanation of that fact. Why a fairly complicated normalized post-Newtonian correction $S_\mathrm{T} $    or
its normalized  compounds:     dragging     $S_\mathrm{g} $,   anti-dragging   $S_\mathrm{ad} $ and centrifugal  $S_\mathrm{c} $ should have  maximal values that almost linearly depend on  the mass functional $M_\mathrm{D}$?
 This scaling would mean that the Dopplerian width of spectral lines, of general-relativistic origin,  emitted by sources corotating with 
Keplerian stationary disks scales proportionately to $M_\mathrm{D}/m$. This opens, in principle at least, a new observational  method for estimating masses in such objects. 

\textit{The mathematical   problems related to stationary rotating polytropes are known as free boundary problems.   There are numerical approaches that might inspire the future mathematics of such systems; we should mention here the pioneering  work of  Hachisu \cite{Hachisu}, Eriguchi and Nishida
 \cite{nishida2} and others \cite{HEAM}.   They are analysed --- with emphasis on the convergence of the SCF approaches --- in the recent work of Price, Markakis and Friedman \cite{PMF}. Hachisu \cite{Hachisu} pointed out the necessity to include the maximal value of the fluid mass density, for rotating newtonian polytropes, in the catalogue of assumed data for the self-consistent field method.   Our work brings a new technical element --- that 
 the maximal (baryonic) mass density should be a part of given data  (at least up to the 1PN order), in addition to the rotation law, the equation of state and information on spatial extendibility. Little is known about mathematical setting of rotating selfgravitating systems within general relativity.}

Finally, we confirmed the validity of estimates formulated by S. Dain \cite{Dain}.  They imply, in particular, that the angular momentum is located mostly in peripherals of rotating black-hole-toroidal systems; this is consistent with the numerical results reported above. 

\begin{acknowledgments}
N.\ Xie is partially supported by the National Science Foundation of China No.11421061. This research was carried out with the supercomputer ``Deszno'' purchased thanks to the financial support of the European Regional Development Fund in the framework of the Polish Innovation Economy Operational Program (contract no.\ POIG.\ 02.01.00-12-023/08).   
\end{acknowledgments}
 

\end{document}